\documentstyle[12pt,a4]{article}
\input epsf
\global\arraycolsep=2pt

\begin{document}

\begin{titlepage}
\begin{flushright}
CERN-TH/96-19\\
SHEP 96-03\\
hep-ph/9603202
\end{flushright}

\vspace{0.5cm}
\begin{center}
\Large\bf Spectator Effects in Inclusive Decays\\
of Beauty Hadrons
\end{center}

\vspace{1.0cm}
\begin{center}
M. Neubert and C.T. Sachrajda$^*$\\
{\sl Theory Division, CERN, CH-1211 Geneva 23, Switzerland}
\end{center}

\vspace{1.2cm}
\begin{abstract}
We present a model-independent study of spectator effects, which are
responsible for the lifetime differences between beauty hadrons.
These effects can be parametrized in terms of hadronic matrix
elements of four four-quark operators. For $B$ mesons, the
coefficients of the non-factorizable operators turn out to be much
larger than those of the factorizable ones, limiting considerably the
usefulness of the vacuum insertion approximation. Non-factorizable
contributions to the lifetime ratio $\tau(B^-)/\tau(B_d)$ could
naturally be of order 10--20\%, and not even the sign of these
contributions can be predicted at present. In the case of the
$\Lambda_b$ baryon, heavy-quark symmetry is used to reduce the number
of independent matrix elements from four to two. In order to explain
the large deviation from unity in the experimental result for
$\tau(\Lambda_b)/\tau(B_d)$, it is necessary that these baryon matrix
elements be much larger than those estimated in quark models. We have
also reexamined the theoretical predictions for the semileptonic
branching ratio of $B$ mesons and charm counting, finding that, given
the present theoretical and experimental uncertainties, there is no
significant discrepancy with experiment.
\end{abstract}

\vspace{1.0cm}
\centerline{(Revised Version)}

\vspace{2.0cm}
\noindent
CERN-TH/96-19\\
September 1996

\vspace{1.0cm}
\centerline{$^*$\small
On leave from the Department of Physics, University of Southampton,
Southampton SO17 1BJ, UK}

\end{titlepage}

\section{Introduction}
\label{sec:intro}

In this paper we study ``spectator effects'' in inclusive decays of
beauty hadrons. These effects involve the participation of the light
constituents in the decay and thus contribute to the differences in
the decay widths and lifetimes of different species of beauty
hadrons. Indeed, one of our goals is to understand theoretically the
experimental results for the lifetime ratios \cite{Joe}:
\begin{eqnarray}\label{taudata}
   {\tau(B^-)\over\tau(B_d)} &=& 1.02\pm 0.04 \,, \nonumber\\
   {\tau(B_s)\over\tau(B_d)} &=& 1.01\pm 0.07 \,, \nonumber\\
   {\tau(\Lambda_b)\over\tau(B_d)} &=& 0.78 \pm 0.05 \,.
\end{eqnarray}
Here $\tau(B_s)$ refers to the average $B_s$-meson lifetime. Our
study is performed in the framework of the heavy-quark expansion, in
which these ratios are computed as series in inverse powers of the
mass of the $b$ quark \cite{Chay}--\cite{Blok} (for recent reviews,
see refs.~\cite{liferef,Beijing}). The leading term of this expansion
corresponds to the decay of a free $b$ quark. This term is universal,
contributing equally to the lifetimes of all beauty hadrons.
Remarkably, the first correction to this result is of order
$(\Lambda_{\rm QCD}/m_b)^2$ \cite{Bigi,MaWe}. This leads to the
theoretical predictions (see section~\ref{sec:ope} below)
\begin{eqnarray}\label{taucrude}
   {\tau(B^-)\over\tau(B_d)} &=& 1 + O(1/m_b^3) \,,
    \nonumber\\
   {\tau(B_s)\over\tau(B_d)} &=& (1.00\pm 0.01) + O(1/m_b^3) \,,
    \nonumber\\
   {\tau(\Lambda_b)\over\tau(B_d)} &=& 0.98 + O(1/m_b^3) \,.
\end{eqnarray}
The deviation from this expectation for $\tau(\Lambda_b)/\tau(B_d)$
is striking, and is the principle motivation for this study.

Spectator effects, i.e.\ contributions from decays in which a light
constituent quark also participates in the weak process, have first
been considered in refs.~\cite{Gube}--\cite{ShiV}. For decays of
heavy particles, these effects are strongly suppressed due to the
need for the $b$ quark and a light quark in the heavy hadron to be
close together (i.e. from a factor of the ``wave-function at the
origin''). As the portion of the volume that the $b$ quark occupies
inside the hadron is of order $(\Lambda_{\rm QCD}/m_b)^3$, spectator
effects appear only at third order in the heavy-quark expansion, and
it might seem safe to neglect them altogether. However, as a result
of the difference in the phase-space for $2\to 2$-body reactions as
compared to $1\to 3$-body decays, these effects are enhanced by a
factor of order $16\pi^2$. It is conceivable that they
could be larger than the terms of order $(\Lambda_{\rm QCD}/m_b)^2$
included in (\ref{taucrude}). Moreover, spectator effects explicitly
differentiate between different species of beauty hadrons. In order
to understand the structure of lifetime differences, it is therefore
important to reconsider the analysis of such effects. The striking
experimental result for the short $\Lambda_b$ lifetime gives an
additional motivation to such a study.

Previous studies of spectator effects in the decays of beauty hadrons
\cite{liferef} were performed using the formalism developed in
refs.~\cite{Gube}--\cite{ShiV} and made two simplifying assumptions:
first, the hadronic matrix elements of these operators were estimated
employing the vacuum insertion approximation \cite{SVZ} for mesons
and quark models \cite{Gube,Cort} for baryons, and secondly the mass
of the charm quark was neglected in the calculation of the
coefficients of the four-quark operators in the heavy-quark
expansion. In order to explore the acceptable range of theoretical
predictions, we do not impose factorization or quark-model
approximations on the hadronic matrix elements. Instead, we
parametrize them by a set of hadronic parameters and see how the
lifetime ratios depend upon them. One of our main conclusions is that
only a detailed field-theoretic calculation of the relevant matrix
elements can lead to reliable predictions. In addition, we derive the
exact expressions for the coefficients as functions of the
charm-quark mass.

The semileptonic branching ratio of $B$ mesons has also received
considerable attention. For many years it appeared that the
theoretical predictions for this quantity \cite{Cort1}--\cite{baff}
lay above the measured value. More recently, the theoretical
predictions have been refined by including exact expressions for the
$O(\alpha_s)$ corrections \cite{Baga}. Using the results of these
calculations, we present a new analysis of the semileptonic branching
ratio $B_{\rm SL}$ and the average charm multiplicity $n_c$ in $B$
decays. We find that the freedom in the choice of the renormalization
scale (which reflects the ignorance of higher-order perturbative
corrections) allows us to obtain consistent predictions for both
quantities simultaneously. We also calculate the spectator
contributions to $B_{\rm SL}$ and $n_c$ and show that they could
change the semileptonic branching ratio by an amount of order 1\%,
whereas their effect on $n_c$ is negligible.

In section~\ref{sec:ope}, we discuss the heavy-quark expansion for
inclusive decay rates, and we present our results for the
contributions arising from spectator effects. In
section~\ref{sec:param}, we introduce a set of hadronic parameters
defined in terms of the relevant four-quark operator matrix elements
between $B$-meson and $\Lambda_b$-baryon states. Heavy-quark symmetry
is used to derive some new relations between the baryonic matrix
elements of these operators. In section~\ref{sec:phen}, we then
discuss the phenomenological implications of our results for the
understanding of beauty lifetimes. We also present a critical
discussion of previous estimates of spectator effects based on the
factorization approximation. A detailed discussion of the
semileptonic branching ratio of $B$ mesons is presented in
section~\ref{sec:Bsl}. Section~\ref{sec:concs} contains the
conclusions. The renormalization of the operators and parameters
describing the spectator effects, and the behaviour of these
parameters with respect to the large-$N_c$ limit, are discussed in
appendix~A, while appendix~B contains details about the calculation
of the semileptonic branching ratio.

The reader who is primarily interested in the phenomenological
implications of our analysis can omit sections~\ref{sec:ope} and
\ref{sec:param} and proceed directly to sections~\ref{sec:phen},
\ref{sec:Bsl} and \ref{sec:concs}, which are written in a
self-contained way.

\section{Heavy-quark expansion}
\label{sec:ope}

Inclusive decay rates, which determine the probability of the decay
of a particle into the sum of all possible final states with a given
set of quantum numbers $\{f\}$, have two advantages from the
theoretical point of view: first, bound-state effects related to
the initial state can be accounted for in a systematic way using the
heavy-quark expansion; secondly, the fact that the final state
consists of a sum over many hadronic channels eliminates bound-state
effects related to the properties of individual hadrons. This second
feature is based on the hypothesis of quark--hadron duality, i.e.\
the assumption that cross sections and decay rates are calculable in
QCD after a ``smearing'' procedure has been applied \cite{PQW}. We
shall not discuss this hypothesis here; however, if after the
non-perturbative evaluation of the spectator effects discussed in our
analysis there remain significant discrepancies between theory and
experiment (for the lifetime ratio $\tau(\Lambda_b)/\tau(B_d)$, in
particular), one may have to seriously question the assumption of
duality. A recent study of inclusive $B$ decays, in which duality
violations are invoked to add non-perturbative contributions of order
$\Lambda_{\rm QCD}/m_b$ not present in the heavy-quark expansion, can
be found in ref.~\cite{guido}.

Using the optical theorem, the inclusive decay width of a hadron
$H_b$ containing a $b$ quark can be written as the forward matrix
element of the imaginary part of the transition operator ${\bf T}$,
\begin{equation}\label{ImT}
   \Gamma(H_b\to X) = {1\over m_{H_b}}\,\mbox{Im}\,
   \langle H_b|\,{\bf T}\,|H_b\rangle = {1\over 2 m_{H_b}}\,
   \langle H_b|\,{\bf\Gamma}\,|H_b\rangle \,,
\end{equation}
where ${\bf T}$ is given by
\begin{equation}\label{Top}
   {\bf T} = i\!\int{\rm d}^4 x\,T\{\,
   {\cal L}_{\rm eff}(x),{\cal L}_{\rm eff}(0)\,\} \,.
\end{equation}
For the case of semileptonic and non-leptonic decays, the effective
weak Lagrangian, renormalized at the scale $\mu=m_b$, is
\begin{eqnarray}
   {\cal L}_{\rm eff} &=& - {4 G_F\over\sqrt{2}}\,V_{cb}\,
    \bigg\{ c_1(m_b)\,\Big[
    \bar d'_L\gamma_\mu u_L\,\bar c_L\gamma^\mu b_L +
    \bar s'_L\gamma_\mu c_L\,\bar c_L\gamma^\mu b_L \Big] \nonumber\\
   &&\phantom{ - {4 G_F\over\sqrt{2}}\,V_{cb}\, }
    \mbox{}+ c_2(m_b)\,\Big[
    \bar c_L\gamma_\mu u_L\,\bar d'_L\gamma^\mu b_L +
    \bar c_L\gamma_\mu c_L\,\bar s'_L\gamma^\mu b_L \Big] \nonumber\\
   &&\phantom{ - {4 G_F\over\sqrt{2}}\,V_{cb}\, }
    \mbox{}+ \sum_{\ell=e,\mu,\tau}
    \bar\ell_L\gamma_\mu\nu_\ell\,\bar c_L\gamma^\mu b_L
    \bigg\} + \mbox{h.c.} \,,
\end{eqnarray}
where $q_L=\frac{1}{2}(1-\gamma_5) q$ denotes a left-handed quark
field, $d'=d\,\cos\theta_c + s\,\sin\theta_c$ and $s'=s\,\cos\theta_c
- d\,\sin\theta_c$ are the Cabibbo-rotated down- and strange-quark
fields ($\sin\theta_c\simeq 0.2205$), and we have neglected $b\to u$
transitions. The Wilson coefficients $c_1$ and $c_2$ take into
account the QCD corrections arising from the fact that the effective
Lagrangian is written at a renormalization scale $\mu=m_b$ rather
than $m_W$. They can be calculated in perturbation theory. The
combinations $c_\pm = c_1\pm c_2$ have a multiplicative evolution
under change of the renormalization scale. To leading order, they are
given by \cite{cpcm1}--\cite{cpcm3}
\begin{equation}\label{cpm}
   c_\pm(m_b) = \left( {\alpha_s(m_W)\over\alpha_s(m_b)}
   \right)^{a_\pm} \,,\qquad
   a_- = -2 a_+ = - {12\over 33 - 2 n_f} \,.
\end{equation}
In the numerical analysis we shall take the values $c_+(m_b)\simeq
0.86$ and $c_-(m_b)=1/c_+^2(m_b)\simeq 1.35$, corresponding to
$\alpha_s(m_Z)=0.117$.

Since the energy release in the decay of a $b$ quark is large, it is
possible to construct an Operator Product Expansion (OPE) for the
bilocal transition operator (\ref{Top}), in which it is expanded as a
series of local operators with increasing dimension, whose
coefficients contain inverse powers of the $b$-quark mass. The
operator with the lowest dimension is $\bar b b$. There is no
independent operator with dimension four, since the only candidate,
$\bar b\,i\rlap{\,/}D\,b$, can be reduced to $\bar b b$ by using the
equations of motion \cite{Bigi,MaWe}. The first new operator is $\bar
b\,g_s\sigma_{\mu\nu} G^{\mu\nu} b$ and has dimension five. Thus, any
inclusive decay rate of a hadron $H_b$ can be written in the form
\begin{equation}\label{gener}
   \Gamma(H_b\to X_f) = {G_F^2 m_b^5\over 192\pi^3}\,
   {1\over 2 m_{H_b}}\,\Bigg\{
   c_3^f\,\langle H_b|\,\bar b b\,|H_b\rangle
   + c_5^f\,{\langle H_b|\,\bar b\,g_s\sigma_{\mu\nu} G^{\mu\nu}
   b\,|H_b\rangle\over m_b^2} + \dots \Bigg\} \,,
\end{equation}
where $c_n^f$ are calculable coefficient functions (which also
contain the relevant CKM matrix elements) depending on the quantum
numbers $\{f\}$ of the final state. For semileptonic and
non-leptonic decays, the coefficients $c_3^f$ have been calculated at
one-loop order \cite{Hoki,Nir,Baga}, and the coefficients $c_5^f$
at tree level \cite{Bigi,FLNN}.

In the next step, the forward matrix elements of the local operators
in the OPE are systematically expanded in inverse powers of the
$b$-quark mass, using the heavy-quark effective theory (HQET)
\cite{review}. One finds \cite{Bigi,MaWe}
\begin{eqnarray}\label{mumu}
   {1\over 2 m_{H_b}}\,\langle H_b|\,\bar b b\,|H_b\rangle
   &=& 1 - {\mu_\pi^2(H_b)-\mu_G^2(H_b)\over 2 m_b^2}
    + O(1/m_b^3) \,, \nonumber\\
   {1\over 2 m_{H_b}}\,\langle H_b|\,\bar b\,g_s\sigma_{\mu\nu}
   G^{\mu\nu} b\,|H_b\rangle &=& 2\mu_G^2(H_b) + O(1/m_b) \,,
\end{eqnarray}
where $\mu_\pi^2(H_b)$ and $\mu_G^2(H_b)$ parametrize the matrix
elements of the kinetic-energy and the chromo-magnetic operators,
respectively. The purpose of doing this expansion is that whereas the
matrix elements in (\ref{gener}) contain an implicit dependence on
the $b$-quark mass, the parameters appearing on the right-hand side
of (\ref{mumu}) are independent of $m_b$ (modulo logarithms). These
parameters can be determined, to some extent, from the spectrum of
heavy hadron states. Below we shall need the values
\begin{eqnarray}\label{muvals}
   \mu_\pi^2(\Lambda_b) - \mu_\pi^2(B) &=&
    - (0.01\pm 0.03)~\mbox{GeV}^2 \,, \nonumber\\
   \mu_G^2(B) &=& {3\over 4}\,(m_{B^*}^2 - m_B^2)
    \simeq 0.36~\mbox{GeV}^2 \,, \nonumber\\
   \mu_G^2(\Lambda_b) &=& 0 \,.
\end{eqnarray}
The difference $\mu_\pi^2(\Lambda_b)-\mu_\pi^2(B)$ can be extracted
from the mass formula \cite{liferef,Beijing}
\begin{equation}
   (m_{\Lambda_b}-m_{\Lambda_c}) - (\overline{m}_B-\overline{m}_D)
   = \Big[ \mu_\pi^2(B)-\mu_\pi^2(\Lambda_b) \Big]\,\bigg(
   {1\over 2 m_c} - {1\over 2 m_b} \bigg) + O(1/m_Q^2) \,,
\end{equation}
where $\overline{m}_B=\frac{1}{4}\,(m_B + 3 m_{B^*})$ and
$\overline{m}_D=\frac{1}{4}\,(m_D + 3 m_{D^*})$ denote the
spin-averaged meson masses. With $m_{\Lambda_b}=(5625\pm 6)$~MeV
\cite{mLambda}, this relation leads to the value quoted above.

To order $1/m_b^2$ in the heavy-quark expansion, the lifetime ratio
for two beauty hadrons is given by
\begin{equation}\label{taurat}
   {\tau(H_b^{(1)})\over\tau(H_b^{(2)})} = 1
   + {\mu_\pi^2(H_b^{(1)}) - \mu_\pi^2(H_b^{(2)})\over 2 m_b^2}
   + c_G\,{\mu_G^2(H_b^{(1)}) - \mu_G^2(H_b^{(2)})\over m_b^2}
   + O(1/m_b^3) \,,
\end{equation}
where $c_G\simeq 1.2$ can be obtained using the results of
refs.~\cite{Bigi,FLNN}. Using then the values given in
(\ref{muvals}), and assuming that in the case of the $B_s$ meson
SU(3)-breaking effects in the values of the matrix elements are of
order 20\%, we arrive at the predictions given in (\ref{taucrude}).
Note that in taking a ratio of lifetimes, theoretical uncertainties
related to the values of the $b$-quark mass (including renormalon
ambiguities) and CKM elements cancel to a large extent. It is for
this reason that we restrict our discussion to the calculation of
ratios of lifetimes and decay rates.

\begin{figure}[htb]
   \epsfxsize=9cm
   \centerline{\epsffile{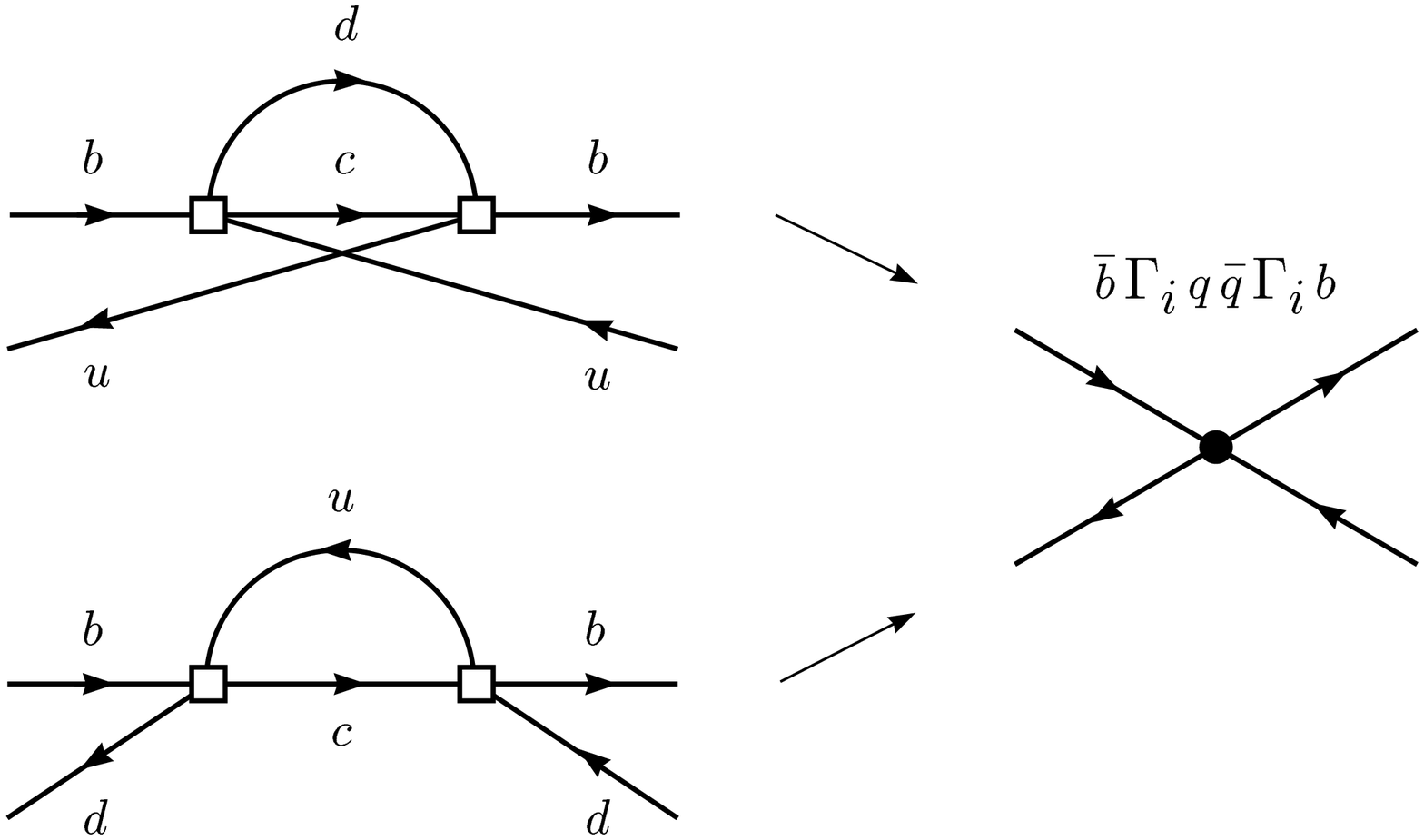}}
   \centerline{\parbox{14cm}{\caption{\label{fig:Tspec}
Spectator contributions to the transition operator ${\bf T}$ (left),
and the corresponding operator in the OPE (right). Here $\Gamma_i$
denotes some combination of Dirac and colour matrices.
   }}}
\end{figure}

The first two terms in the heavy-quark expansion (\ref{gener}) arise
from decays in which the (light) spectator quarks interact only
softly. Additional contributions of this type appear at the next
order through gluonic operators of dimension six, such as $\bar
b\,\gamma_\mu(i D_\nu G^{\mu\nu}) b$. Since matrix elements of these
operators are blind to the flavour of the spectator quarks, they can
be safely neglected in our analysis. ``Hard'' spectator effects
manifest themselves first in the matrix elements of four-quark
operators of dimension six. Some examples of the corresponding
contributions to the transition operator ${\bf T}$ are shown in
figure~\ref{fig:Tspec}. They only appear in the heavy-quark expansion
of non-leptonic decay rates.\footnote{This is no longer true if $b\to
u$ transitions or the decays of the $B_c$ meson are considered.}
Since these contributions arise from one-loop rather than two-loop
diagrams, they receive a phase-space enhancement factor of order
$16\pi^2$ relative to the other terms in the heavy-quark expansion.

We have calculated the coefficients of the corresponding four-quark
operators at tree level, including for the first time the dependence
on the mass of the charm quark. This extends the results obtained in
ref.~\cite{liferef}. We find that the corresponding contributions to
the non-leptonic widths of mesons and baryons containing a $b$ quark
are given by the matrix elements of the local operator
\begin{eqnarray}\label{Gspec}
   {\bf\Gamma}_{\rm spec}
   &=& {2 G_F^2 m_b^2\over\pi}\,|V_{cb}|^2\,(1-z)^2\,\Bigg\{
    \bigg( 2 c_1 c_2 + {1\over N_c}\,(c_1^2 + c_2^2) \bigg)\,
    O_{V-A}^u + 2 (c_1^2 + c_2^2)\,T_{V-A}^u \Bigg\} \nonumber \\
   &-& {2 G_F^2 m_b^2\over 3\pi}\,|V_{cb}|^2\,(1-z)^2\,\Bigg\{
    \bigg( 2 c_1 c_2 + {1\over N_c}\,c_1^2 + N_c\,c_2^2 \bigg)\,
    \bigg[ \bigg( 1 + {z\over 2} \bigg)\,O_{V-A}^{d'}
    - (1+2z)\,O_{S-P}^{d'} \bigg] \nonumber\\
   &&\phantom{ {2 G_F^2 m_b^2\over\pi}\,|V_{cb}|^2\,(1-z)^2\, }
    \mbox{}+ 2 c_1^2\,\bigg[ \bigg( 1 + {z\over 2} \bigg)\,
    T_{V-A}^{d'} - (1+2z)\,T_{S-P}^{d'} \bigg] \Bigg\} \nonumber\\
   &-& {2 G_F^2 m_b^2\over 3\pi}\,|V_{cb}|^2\,\sqrt{1-4 z}\,\Bigg\{
    \bigg( 2 c_1 c_2 + {1\over N_c}\,c_1^2 + N_c\,c_2^2 \bigg)\,
    \Big[ (1-z)\,O_{V-A}^{s'} - (1+2z)\,O_{S-P}^{s'} \Big]
    \nonumber\\
   &&\phantom{ {2 G_F^2 m_b^2\over\pi}\,|V_{cb}|^2\,(1-z)^2\, }
    \mbox{}+ 2 c_1^2\,\Big[ (1-z)\,T_{V-A}^{s'} - (1+2z)\,
    T_{S-P}^{s'} \Big] \Bigg\} \,,
\end{eqnarray}
where $z=m_c^2/m_b^2$, and $N_c$=3 is the number of colours. The
local four-quark operators appearing in this expression are defined
by
\begin{eqnarray}\label{4qops}
   O_{V-A}^q &=& \bar b_L\gamma_\mu q_L\,\bar q_L\gamma^\mu b_L
    \,, \nonumber\\
   O_{S-P}^q &=& \bar b_R\,q_L\,\bar q_L\,b_R \,, \nonumber\\
   T_{V-A}^q &=& \bar b_L\gamma_\mu t_a q_L\,\bar q_L\gamma^\mu
    t_a b_L \,, \nonumber\\
   T_{S-P}^q &=& \bar b_R\,t_a q_L\,\bar q_L\,t_a b_R \,,
\end{eqnarray}
where $t_a=\lambda_a/2$ are the generators of colour SU(3). For
dimensional reasons, ${\bf\Gamma}_{\rm spec}$ is proportional to
$m_b^2$ rather than $m_b^5$, in accordance with the fact that
spectator effects contribute at third order in the heavy-quark
expansion. The first term in (\ref{Gspec}) arises from the upper
diagram in figure~\ref{fig:Tspec}, whereas the second and third terms
come from the contributions of the lower diagram with a $c\bar u$ and
$c\bar c$ quark pair in the loop. We note that in the limit $z=0$ our
results agree with ref.~\cite{liferef}, and with the corresponding
expression derived for the lifetimes of charm hadrons in
refs.~\cite{Gube}--\cite{ShiV}.

The operators in (\ref{4qops}) are renormalized at the scale $m_b$,
which will be implicit in our discussion below. This choice has the
advantage that logarithms of the type $[\alpha_s\ln(m_b/\mu_{\rm
had})]^n$, where $\mu_{\rm had}$ is a typical hadronic scale, reside
entirely in the hadronic matrix elements of the renormalized
operators. Using the renormalization-group equations, the expressions
presented in this paper can be rewritten in terms of operators
renormalized at any other scale. However, at present the scale
dependence of the renormalized operators below the scale $m_b$ is
known only to leading logarithmic order \cite{hybr,PoWi}. It is
discussed in detail in appendix~A. Since here we shall treat the
matrix elements as unknown parameters, we can avoid all uncertainties
related to the operator evolution by working at the scale $m_b$.

The hadronic matrix elements of the four-quark operators in
(\ref{4qops}) contain the non-perturbative physics of the spectator
contributions to inclusive decays of beauty hadrons. However, the
same operators also contribute to the decay of the $b$ quark, through
tadpole diagrams in which the light-quark fields are contracted in a
loop. These ``non-spectator'' contributions are independent of the
flavour of the light quark $q$ and thus contribute equally to the
decay widths of all beauty hadrons. They are not of interest to our
discussion here. In order to isolate the true spectator effects, we
shall implicitly assume a normal ordering of the four-quark
operators, which has the effect of subtracting tadpole-like diagrams.
In practice, this is equivalent to choosing a particular
renormalization prescription for the operators. Alternatively, one
may isolate the spectator effects by considering light-quark flavour
non-singlet combinations of the operators; thus, for example, for
matrix elements between $B_d$ states one could take $(O^d-O^u)$ and
$(T^d-T^u)$ instead of $O^d$ and $T^d$.

\section{Parametrization of the matrix elements}
\label{sec:param}

In previous analyses \cite{liferef}, \cite{Gube}--\cite{ShiV}, the
hadronic matrix elements of the four-quark operators in (\ref{4qops})
have been estimated making simplifying assumptions. Here we shall
avoid such assumptions and, without any loss of generality, express
the relevant matrix elements in terms of a set of hadronic
parameters. Clearly, such an approach has less predictive power;
however, it does allow us to find the range of predictions that can
be made in a model-independent way. In view of the apparent
discrepancy between theory and experiment for the $\Lambda_b$
lifetime, we find it worth while to question the model-dependent
assumptions made in earlier analyses. Ultimately, the relevant
hadronic parameters may be calculated using some field-theoretic
approach such as lattice gauge theory or QCD sum rules. It has also
been suggested that combinations of these parameters may be extracted
from a precise measurement of the lepton spectrum in the endpoint
region of semileptonic $B$ decays, or from a study of spectator
effects in charm decays \cite{BiUr}.

\subsection{Mesonic matrix elements}

For matrix elements of the four-quark operators between $B$-meson
states, we define parameters $B_i$ and $\varepsilon_i$ such that:
\begin{eqnarray}\label{Bidef}
   {1\over 2 m_{B_q}}\,\langle B_q|\,O_{V-A}^q\,|B_q\rangle
   &\equiv& {f_{B_q}^2 m_{B_q}\over 8}\,B_1 \,, \nonumber\\
   {1\over 2 m_{B_q}}\,\langle B_q|\,O_{S-P}^q\,|B_q\rangle
   &\equiv& {f_{B_q}^2 m_{B_q}\over 8}\,B_2 \,, \nonumber\\
   {1\over 2 m_{B_q}}\,\langle B_q|\,T_{V-A}^q\,|B_q\rangle
   &\equiv& {f_{B_q}^2 m_{B_q}\over 8}\,\varepsilon_1 \,,
    \nonumber\\
   {1\over 2 m_{B_q}}\,\langle B_q|\,T_{S-P}^q\,|B_q\rangle
   &\equiv& {f_{B_q}^2 m_{B_q}\over 8}\,\varepsilon_2 \,.
\end{eqnarray}
This definition is inspired by the vacuum insertion (or
factorization) approximation \cite{SVZ}, according to which the
matrix elements of four-quark operators are evaluated by inserting
the vacuum inside the current products. This leads to
\begin{eqnarray}\label{fact}
   \langle B_q|\,O_{V-A}^q\,|B_q\rangle
   &=& \bigg( {m_b+m_q\over m_B} \bigg)^2\,
    \langle B_q|\,O_{S-P}^q\,|B_q\rangle
    = {f_{B_q}^2 m_{B_q}^2\over 4} \,, \nonumber\\
   \langle B_q|\,T_{V-A}^q\,|B_q\rangle
   &=& \langle B_q|\,T_{S-P}^q\,|B_q\rangle = 0 \,,
\end{eqnarray}
where $f_{B_q}$ is the decay constant of the $B_q$ meson, defined as
\begin{equation}
   \langle 0\,|\,\bar q\,\gamma^\mu\gamma_5\,b\,|B_q(p)\rangle
   = i f_{B_q}\,p^\mu \,.
\end{equation}
Hence, the factorization approximation corresponds to setting $B_i=1$
and $\varepsilon_i=0$ at some scale $\mu$ (which in general will be
different from our adopted choice $\mu=m_b$), where the approximation
is believed to be valid. The exact values of the hadronic parameters
are not yet known. However, as discussed in appendix~A, in the
large-$N_c$ limit
\begin{equation}
   B_i = O(1) \,,\qquad \varepsilon_i=O(1/N_c) \,.
\end{equation}
An estimate of the parameters $\varepsilon_i$ using QCD sum rules has
been obtained by Chernyak, who finds that $\varepsilon_1\approx
-0.15$ and $\varepsilon_2\approx 0$ \cite{Chern}.

In terms of the parameters $B_i$ and $\varepsilon_i$, the matrix
elements of the operator ${\bf\Gamma}_{\rm spec}$ in (\ref{Gspec})
are:
\begin{eqnarray}\label{mesons}
\lefteqn{
   {1\over 2 m_B}\,\langle B^-|\,{\bf\Gamma}_{\rm spec}\,
   |B^-\rangle = \Gamma_0\,\eta_{\rm spec}\,(1-z)^2\,\Big\{
    (2 c_+^2 - c_-^2)\,B_1 + 3 (c_+^2 + c_-^2)\,\varepsilon_1 \Big\}
    \,, } \nonumber\\
\lefteqn{
   {1\over 2 m_B}\,\langle B_d|\,{\bf\Gamma}_{\rm spec}\,|B_d\rangle
   } \nonumber\\
    &&= - \Gamma_0\,\eta_{\rm spec}\,(1-z)^2\,\cos^2\!\theta_c\,
    \Bigg\{ {1\over 3}\,(2 c_+ - c_-)^2\,\bigg[
    \bigg( 1 + {z\over 2} \bigg)\,B_1 - (1+2z)\,B_2 \bigg]
    \nonumber\\
   &&\quad\mbox{}+ {1\over 2}\,(c_+ + c_-)^2\,\bigg[
    \bigg( 1 + {z\over 2} \bigg)\,\varepsilon_1
    - (1+2z)\,\varepsilon_2 \bigg] \Bigg\} \nonumber\\
   &&\quad\mbox{}- \Gamma_0\,\eta_{\rm spec}\,\sqrt{1-4z}\,
    \sin^2\!\theta_c\,\Bigg\{ {1\over 3}\,(2 c_+ - c_-)^2\,\Big[
    (1-z)\,B_1 - (1+2z)\,B_2 \Big] \nonumber\\
   &&\quad\mbox{}+ {1\over 2}\,(c_+ + c_-)^2\,\Big[
    (1-z)\,\varepsilon_1 - (1+2z)\,\varepsilon_2 \Big] \Bigg\} \,,
\end{eqnarray}
where $c_\pm=c_1\pm c_2$, and
\begin{equation}\label{Gamma0}
   \Gamma_0 = {G_F^2 m_b^5\over 192 \pi^3}\,|V_{cb}|^2 \,,\qquad
   \eta_{\rm spec} = 16\pi^2\,{f_B^2 m_B\over m_b^3} \,.
\end{equation}
The spectator contribution to the width of the $B_s$ meson is
obtained from that of the $B_d$ meson by the replacements
$\sin\theta_c\leftrightarrow \cos\theta_c$ and $f_B,m_B\to
f_{B_s},m_{B_s}$. Of course, the values of the parameters $B_i$ and
$\varepsilon_i$ for the $B_s$ meson will also differ from those for
the $B_d$ meson due to SU(3)-breaking effects.

Two remarks are in order regarding the result (\ref{mesons}). The
first concerns the expected order of magnitude of the spectator
contributions to the total decay width of a $B$ meson. At leading
order in the heavy-quark expansion, the total width of a beauty
hadron is $\Gamma_{\rm tot}\simeq 3.7\times\Gamma_0$, where the
numerical factor arises from the phase-space contributions of the
semileptonic and non-leptonic channels (we use $z=0.085$)
\cite{Bigi,FLNN}. It follows that
\begin{equation}
   {\Gamma_{\rm spec}\over\Gamma_{\rm tot}}
   \sim {\eta_{\rm spec}\over 4}
   \sim \bigg( {2\pi f_B\over m_B} \bigg)^2 \sim 5\% \,.
\end{equation}
The second remark concerns the structure of the coefficients in
(\ref{mesons}). Given that $c_+\simeq 0.86$ and $c_-\simeq 1.35$, one
observes that the coefficients of the colour singlet--singlet
operators are one to two orders of magnitude smaller than those of
the colour octet--octet operators. This implies that at the scale
$m_b$ even small deviations from the factorization approximation can
have a sizeable impact on the results.

\subsection{Baryonic matrix elements}
\label{subs:baryon}

Next we study the matrix elements of the four-quark operators between
$\Lambda_b$-baryon states. Since the $\Lambda_b$ is an iso-singlet,
the matrix elements of the operators with $q=u$ or $d$ are the same,
and below we drop this label. In the case of baryons, we find it
convenient to use the colour identity
\begin{equation}
   (t_a)_{\alpha\beta}\,(t_a)_{\gamma\delta}
   = {1\over 2}\,\delta_{\alpha\delta}\,\delta_{\gamma\beta}
   - {1\over 2 N_c}\,\delta_{\alpha\beta}\,\delta_{\gamma\delta}
\end{equation}
to rewrite $T=-\frac{1}{6}\,O + \frac{1}{2}\,\widetilde O$ and
introduce the operators ($i,j$ are colour indices)
\begin{equation}
   \widetilde O_{V-A} = \bar b_L^i\gamma_\mu q_L^j\,
    \bar q_L^j\gamma^\mu b_L^i \,, \qquad
   \widetilde O_{S-P} = \bar b_R^i\,q_L^j\,\bar q_L^j\,b_R^i \,.
\end{equation}
instead of $T_{V-A}$ and $T_{S-P}$.

The heavy-quark spin symmetry, i.e.\ the fact that interactions with
the spin of the heavy quark decouple as the heavy-quark mass tends to
infinity, allows us to derive two relations between the matrix
elements of the four-quark operators between $\Lambda_b$-baryon
states. To find these relations, we note that the following matrix
element vanishes in the limit $m_b\to\infty$:
\begin{equation}\label{axial}
   {1\over 2 m_{\Lambda_b}}\,\langle\Lambda_b|\,\bar b^i
   \gamma_\mu\gamma_5\,b^j\,\bar q_L^k\gamma^\mu q_L^l\,
   |\Lambda_b\rangle = O(1/m_b) \,.
\end{equation}
The physical argument for this is that, because of the spin symmetry
for heavy quarks, the matrix elements for left-handed and
right-handed $b$ quarks must be the same. The above result then
follows since $\bar b\,\gamma_\mu\gamma_5\,b = \bar b_R\gamma_\mu b_R
- b_L\gamma_\mu b_L$. A more formal argument can be given using the
covariant tensor formalism of the HQET \cite{review,FGGW} to show
that the matrix element in (\ref{axial}) is proportional to
\begin{equation}
   \langle 2 {\bf S}_b\cdot {\bf S}_{\rm light} \rangle
   = S_{\Lambda_b} (S_{\Lambda_b}+1) - S_b (S_b+1)
   - S_{\rm light} (S_{\rm light}+1) = 0 \,,
\end{equation}
since, in the heavy-quark limit, the light degrees of freedom are in
a state with total spin zero.

Using the Fierz identity
\begin{equation}
   \bar b^i\gamma_\mu\gamma_5\,b^j\,\bar q_L^k\gamma^\mu q_L^l
   = -2\,\bar b_R^i\,q_L^l\,\bar q_L^k\,b_R^j
   - \bar b_L^i\gamma_\mu q_L^l\,\bar q_L^k\gamma^\mu b_L^j \,,
\end{equation}
we then obtain the relations
\begin{eqnarray}
   {1\over 2 m_{\Lambda_b}}\,\langle\Lambda_b|\,O_{S-P}\,
   |\Lambda_b\rangle &=& - {1\over 2}\,{1\over 2 m_{\Lambda_b}}\,
    \langle\Lambda_b|\,O_{V-A}\,|\Lambda_b\rangle
    + O(1/m_b) \,, \nonumber\\
   {1\over 2 m_{\Lambda_b}}\,\langle\Lambda_b|\,\widetilde O_{S-P}\,
   |\Lambda_b\rangle &=& - {1\over 2}\,{1\over 2 m_{\Lambda_b}}\,
    \langle\Lambda_b|\,\widetilde O_{V-A}\,|\Lambda_b\rangle
    + O(1/m_b) \,.
\end{eqnarray}
The corrections of order $1/m_b$ to these relations contribute at
order $1/m_b^4$ in the heavy-quark expansion and so are negligible to
the order we work in. This leaves us with two independent matrix
elements of the operators $O_{V-A}$ and $\widetilde O_{V-A}$. The
analogue of the factorization approximation in the case of baryons
is the valence-quark assumption, in which the colour of the quark
fields in the operators is identified with the colour of the quarks
inside the baryon. Since the colour wave function for a baryon is
totally antisymmetric, the matrix elements of $O_{V-A}$ and
$\widetilde O_{V-A}$ differ in this approximation only by a sign.
Hence, we define a parameter $\widetilde B$ by
\begin{equation}\label{Btildef}
   \langle\Lambda_b|\,\widetilde O_{V-A}\,|\Lambda_b\rangle \equiv
   - \widetilde B\,\langle\Lambda_b|\,O_{V-A}\,|\Lambda_b\rangle \,,
\end{equation}
with $\widetilde B=1$ in the valence-quark approximation.

For the baryon matrix element of $O_{V-A}$ itself, our
parametrization is guided by the quark model. We write
\begin{equation}\label{rdef}
   {1\over 2 m_{\Lambda_b}}\,\langle\Lambda_b|\,O_{V-A}\,
   |\Lambda_b\rangle \equiv - {f_B^2 m_B\over 48}\,r \,,
\end{equation}
where in the quark model $r$ is the ratio of the squares of the wave
functions determining the probability to find a light quark at the
location of the $b$ quark inside the $\Lambda_b$ baryon and the $B$
meson, i.e.\ \cite{Bili,ShiV}
\begin{equation}
   r = {|\psi_{bq}^{\Lambda_b}(0)|^2\over|\psi_{b\bar q}^{B_q}(0)|^2}
   \,.
\end{equation}
Guberina et al.\ have estimated the ratio $r$ for charm decays and
find $r\simeq 0.2$ in the bag model, and $r\simeq 0.5$ in the
non-relativistic quark model \cite{Gube}. This latter estimate has
been used in more recent work on beauty decays \cite{liferef}. A
similar result, $r\sim 0.1$--0.3, has been obtained by Colangelo and
Fazio using QCD sum rules \cite{Cola}. Recently, Rosner has
criticized the existing quark-model estimates of $r$. Assuming that
the wave functions of the $\Lambda_b$ and $\Sigma_b$ baryons are the
same, he argues that the wave-function ratio should be estimated from
the ratio of the spin splittings between $\Sigma_b$ and $\Sigma_b^*$
baryons and $B$ and $B^*$ mesons \cite{Rosn}. This leads to
\begin{equation}
   r = {4\over 3}\,{m_{\Sigma_b^*}^2 - m_{\Sigma_b}^2\over
                    m_{B^*}^2 - m_B^2} \,.
\end{equation}
If the baryon splitting is taken to be $m_{\Sigma_b^*}^2 -
m_{\Sigma_b}^2\simeq m_{\Sigma_c^*}^2 - m_{\Sigma_c}^2 = (0.384\pm
0.035)~\mbox{GeV}^2$, this leads to $r\simeq 0.9\pm 0.1$. If, on the
other hand, one uses the preliminary result $m_{\Sigma_b^*} -
m_{\Sigma_b} = (56\pm 16)$~MeV reported by the DELPHI Collaboration
\cite{DELPHI}, one obtains $r\simeq 1.8\pm 0.5$. We conclude that it
is conceivable that $r\sim 1$, i.e.\ larger than previous estimates.

In terms of these parameters, the matrix element of ${\bf\Gamma}_{\rm
spec}$ is given by
\begin{eqnarray}\label{baryon}
   {1\over 2 m_{\Lambda_b}}\,\langle\Lambda_b|\,
   {\bf\Gamma}_{\rm spec}\,|\Lambda_b\rangle
   &=& \Gamma_0\,\eta_{\rm spec}\,{r\over 16}\,\Bigg\{
    4 (1-z)^2\,\Big[ (c_-^2 - c_+^2)
    + (c_-^2 + c_+^2)\,\widetilde B \Big] \nonumber\\
   &&\mbox{}- \Big[ (1-z)^2 (1+z)\,\cos^2\!\theta_c
    + \sqrt{1-4 z}\,\sin^2\!\theta_c \Big] \nonumber\\
   &&\quad\times \Big[ (c_- - c_+)(5 c_+ - c_-)
    + (c_- + c_+)^2\,\widetilde B \Big] \Bigg\} \,.
\end{eqnarray}

\subsection{Numerical results}

To illustrate the main features of our results, we calculate the
coefficients of the hadronic parameters $B_i$, $\varepsilon_i$, and
$r$ and $\widetilde B\,r$ in the matrix elements (\ref{mesons}) and
(\ref{baryon}) in units of $\Gamma_0\,\eta_{\rm spec}$. In order to
study the dependence on the mass ratio $z=(m_c/m_b)^2$, we first keep
the values of the Wilson coefficients in the effective Lagrangian
fixed and vary the mass ratio in the range $z=0.085\pm 0.015$. This
leads to the numbers shown in table~\ref{tab:1}, where the variation
with $z$ is indicated as a change in the last digit(s). Note that for
mesons the coefficients of the parameters $B_i$ are much smaller than
those of the parameters $\varepsilon_i$. It is apparent that the
results are rather stable with respect to the precise value of $z$.
{}From now on we shall always use the central value $z=0.085$, which
is obtained, for instance, for $m_c=1.4$~GeV and $m_b=4.8$~GeV.

\begin{table}[htb]
\centerline{\parbox{14cm}{\caption{\label{tab:1}
Coefficients of the hadronic parameters obtained for $z=0.085\pm
0.015$. The values $c_+=0.861$ and $c_-=1.349$ are kept fixed.}}}
\vspace{0.5cm}
\centerline{\begin{tabular}{|c|rrrr|rr|}\hline\hline
\rule[-0.2cm]{0cm}{0.65cm} $H_b$ & \multicolumn{1}{c}{$B_1$} &
 \multicolumn{1}{c}{$B_2$} & \multicolumn{1}{c}{$\varepsilon_1$} &
 \multicolumn{1}{c|}{$\varepsilon_2$} & \multicolumn{1}{c}{$r$} &
 \multicolumn{1}{c|}{$\widetilde B\,r$} \\
\hline
$B^-$ & $-0.28(1)$ & \multicolumn{1}{c}{---} & 6.43(21) &
 \multicolumn{1}{c|}{---} & & \\
$B_d$ & $-0.04(0)$ & 0.05(0) & $-2.12(6)~$ & 2.39(2) & & \\
$B_s$ & $-0.03(0)$ & 0.04(0) & $-1.83(11)$ & 2.32(5) & & \\
$\Lambda_b$ & & & & & 0.14(1) & 0.26(1) \\
\hline\hline
\end{tabular}}
\end{table}

As another check on the stability of the results we present in
table~\ref{tab:2} the coefficients obtained using different
``matching'' procedures. To this end, we renormalize the Wilson
coefficients $c_+$ and $c_-$ of the effective Lagrangian at a scale
$\mu$ different from $m_b$. Thus $c_+$ and $c_-$ are modified by
replacing $m_b$ by $\mu$ in (\ref{cpm}). This gives us predictions in
terms of operators renormalized at $\mu$, which we then rewrite in
terms of those defined at $m_b$ by using the evolution equations
given in appendix~A. Thus, the meaning of the hadronic parameters is
the same as before, and the differences between the numerical results
can be viewed as an estimate of unknown higher-order perturbative
corrections, which we neglect throughout this paper. The coefficients
of the parameters $B_i$ for mesons, as well as of the parameters $r$
and $\widetilde B\,r$ for the $\Lambda_b$ baryon, show a significant
scale dependence. To reduce this dependence would require a full
next-to-leading order calculation of radiative corrections, which is
beyond the scope of this paper.

\begin{table}[htb]
\centerline{\parbox{14cm}{\caption{\label{tab:2}
Coefficients of the hadronic parameters obtained for the matching
scales $\mu=m_b/2$, $m_b$ and $2 m_b$, with $m_b=4.8$~GeV. The value
$z=0.085$ is kept fixed.}}}
\vspace{0.5cm}
\centerline{\begin{tabular}{|c|c|rrrr|rr|}\hline\hline
\rule[-0.2cm]{0cm}{0.65cm} $H_b$ & $\mu$ & \multicolumn{1}{c}{$B_1$}
 & \multicolumn{1}{c}{$B_2$} & \multicolumn{1}{c}{$\varepsilon_1$} &
 \multicolumn{1}{c|}{$\varepsilon_2$} & \multicolumn{1}{c}{$r$} &
 \multicolumn{1}{c|}{$\widetilde B\,r$} \\
\hline
    & $m_b/2$ & $-0.62$ & \multicolumn{1}{c}{---} & 7.26 &
 \multicolumn{1}{c|}{---} & & \\
$B^-$ & $m_b$ & $-0.28$ & \multicolumn{1}{c}{---} & 6.43 &
 \multicolumn{1}{c|}{---} & & \\
    & $2 m_b$ & $+0.02$ & \multicolumn{1}{c}{---} & 5.90 &
 \multicolumn{1}{c|}{---} & & \\
\hline
    & $m_b/2$ & $-0.04$ & 0.04 & $-2.32$ & 2.61 & & \\
$B_d$ & $m_b$ & $-0.04$ & 0.05 & $-2.12$ & 2.39 & & \\
    & $2 m_b$ & $-0.08$ & 0.09 & $-1.98$ & 2.24 & & \\
\hline
    & $m_b/2$ & $-0.03$ & 0.04 & $-2.00$ & 2.54 & & \\
$B_s$ & $m_b$ & $-0.03$ & 0.04 & $-1.83$ & 2.32 & & \\
    & $2 m_b$ & $-0.07$ & 0.08 & $-1.72$ & 2.18 & & \\
\hline
    & $m_b/2$ & & & & & 0.21 & 0.30 \\
$\Lambda_b$ & $m_b$ & & & & & 0.14 & 0.26 \\
    & $2 m_b$ & & & & & 0.09 & 0.23 \\
\hline\hline
\end{tabular}}
\end{table}

\section{Phenomenology of beauty lifetimes}
\label{sec:phen}

We shall now discuss the phenomenological implications of our results
for the calculation of beauty lifetime ratios. The spectator
contributions to the decay widths of $B_q$ mesons and of the
$\Lambda_b$ baryon are described by a set of hadronic parameters:
$B_{1,2}$ and $\varepsilon_{1,2}$ for mesons, and $r$ and $\widetilde
B$ for baryons. The explicit dependence of the decay rates on these
quantities is shown in (\ref{mesons}) and (\ref{baryon}). The
numerical values of the coefficients multiplying the hadronic
parameters are given in table~\ref{tab:2} for three different choices
of the matching scale $\mu$. For the numerical analysis we need the
value of the parameter $\eta_{\rm spec}$ defined in (\ref{Gamma0}).
We take $f_B=200$~MeV and $m_b=4.8$~GeV, so that $\eta_{\rm
spec}\simeq 0.30$, and absorb the uncertainty in $\eta_{\rm spec}$
into the values of the hadronic parameters.

\subsection{Lifetime ratio $\tau(B^-)/\tau(B_d)$}

We start by discussing the lifetime ratio of the charged and neutral
$B$ mesons. Because of isospin symmetry, the lifetimes of these
states are the same at order $1/m_b^2$ in the heavy-quark expansion,
and differences arise only from spectator effects. If we write
\begin{equation}\label{cidef}
   {\tau(B^-)\over\tau(B_d)} = 1 + k_1 B_1 + k_2 B_2
   + k_3\varepsilon_1 + k_4\varepsilon_2 \,,
\end{equation}
the coefficients $k_i$ take the values shown in table~\ref{tab:3}.
The most striking feature of this result is the large imbalance
between the coefficients of the parameters $B_i$ and $\varepsilon_i$,
which parametrize the matrix elements of colour singlet--singlet and
colour octet--octet operators, respectively. With $\varepsilon_i$ of
order $1/N_c$, it is conceivable that the non-factorizable
contributions actually dominate the result. Thus, without a detailed
calculation of the parameters $\varepsilon_i$ no reliable prediction
can be obtained. In this conclusion we disagree with the authors of
ref.~\cite{liferef}, who use factorization (at a low hadronic scale)
to argue that $\tau(B^-)/\tau(B_d)$ must exceed unity by an amount of
order 5\%. We will return to this in section~\ref{sec:fact} below.

\begin{table}[htb]
\centerline{\parbox{14cm}{\caption{\label{tab:3}
Coefficients $k_i$ appearing in (\protect\ref{cidef}).}}}
\vspace{0.5cm}
\centerline{\begin{tabular}{|c|cccc|}\hline\hline
\rule[-0.2cm]{0cm}{0.65cm} $\mu$ & $k_1$ & $k_2$ & $k_3$ & $k_4$ \\
\hline
$m_b/2$ & $+0.044$ & 0.003 & $-0.735$ & 0.201 \\
$m_b$   & $+0.020$ & 0.004 & $-0.697$ & 0.195 \\
$2 m_b$ & $-0.008$ & 0.007 & $-0.665$ & 0.189 \\
\hline\hline
\end{tabular}}
\end{table}

The experimental value of the lifetime ratio given in (\ref{taudata})
can be employed to constrain a certain combination of the parameters:
\begin{equation}\label{constr}
   \varepsilon_1 = {1\over k_3}\,\Big[ (0.02\pm 0.04)
   - k_1 B_1 - k_2 B_2 - k_4\varepsilon_2 \Big] \,.
\end{equation}
With the values of the coefficients given in table~\ref{tab:3}, this
relation implies
\begin{equation}
   \varepsilon_1\simeq 0.3\,\varepsilon_2 + \delta \,,
\end{equation}
where we expect $|\delta|<0.1$. This becomes a particularly useful
constraint if the parameters $\varepsilon_i$ turn out to be large.
Below, we shall use (\ref{constr}) to eliminate $\varepsilon_1$ from
our predictions for spectator effects.

\subsection{Lifetime ratio $\tau(B_s)/\tau(B_d)$}

In the limit where SU(3)-breaking effects are neglected, the
spectator contributions to the decay widths of $B_s$ and $B_d$ mesons
are too similar to produce an observable lifetime difference (see
table~\ref{tab:2}). The corresponding contributions to the lifetime
ratio $\tau(B_s)/\tau(B_d)$ are of order $10^{-3}$. A precise
prediction for SU(3)-breaking effects is difficult to obtain.
Allowing for 30\% SU(3)-breaking effects in the matrix elements of
the four-quark operators describing the spectator contributions, we
estimate that the resulting contributions to the lifetime ratio are
of order 1--2\%. Contributions of order 1\% (or less) could also
arise from SU(3)-breaking effects in the matrix elements appearing at
order $1/m_b^2$ in the expansion (\ref{taurat}). Hence, we agree with
ref.~\cite{liferef} that
\begin{equation}
   {\tau(B_s)\over\tau(B_d)} = 1 \pm O (1\%) \,.
\end{equation}

\subsection{Lifetime ratio $\tau(\Lambda_b)/\tau(B_d)$}
\label{subsec:Lambdab}

As mentioned in the introduction, the low experimental value of the
lifetime ratio $\tau(\Lambda_b)/\tau(B_d)$ is the primary motivation
for our study. We shall now discuss the structure of spectator
contributions to this ratio. It is important that heavy-quark
symmetry allows us to reduce the number of hadronic parameters
contributing to the decay rate of the $\Lambda_b$ baryon from four to
two, and that these parameters are almost certainly positive (unless
the quark model is completely misleading) and enter the decay rate
with the same sign. Thus, unlike in the meson case, the structure of
the spectator contributions to the width of the $\Lambda_b$ baryon is
rather simple, and at least the sign of the effects can be predicted
reliably.

For a more detailed discussion, we distinguish between the two cases
where one does or does not allow spectator contributions to enhance
the theoretical prediction for the semileptonic branching ratio,
$B_{\rm SL}$, of $B$ mesons. As we will discuss below, the
theoretical prediction for $B_{\rm SL}$, which neglects spectator
contributions, is slightly larger than the central experimental
value. If spectator effects increased the prediction for $B_{\rm SL}$
further, this discrepancy could become uncomfortably large.

If we do not allow for an increase in the value of the semileptonic
branching ratio, the explanation of the low value of
$\tau(\Lambda_b)/\tau(B_d)$ must reside entirely in a low value of
the $\Lambda_b$ lifetime (rather than a large value of the $B$-meson
lifetime). This can be seen by writing
\begin{equation}
   {\tau(\Lambda_b)\over\tau(B_d)} = \tau(\Lambda_b)\,
   \left( {\tau(B^-)\over\tau(B_d)} \right)^{1/2}\,
   {1\over\big[\tau(B^-)\,\tau(B_d)\big]^{1/2}} \nonumber\\
   = {\tau(\Lambda_b)\over B_{\rm SL}}\,\left(
   {\tau(B^-)\over\tau(B_d)} \right)^{1/2}\,\Gamma_{\rm SL}(B) \,,
\end{equation}
where $B_{\rm SL}$ is the average semileptonic branching ratio of
$B$ mesons, and $\Gamma_{\rm SL}(B)$ is the semileptonic width. In
the second step we have replaced the geometric mean
$[\tau(B^-)\,\tau(B_d)]^{1/2}$ by the average $B$-meson lifetime,
which because of isospin symmetry is correct to order $1/m_b^6$ in
the heavy-quark expansion. Since there are no spectator contributions
to the semileptonic rate $\Gamma_{\rm SL}(B)$, and since we do not
allow an enhancement of the semileptonic branching ratio, in order
to obtain a small value for $\tau(\Lambda_b)/\tau(B_d)$ we can
increase the width of the $\Lambda_b$ baryon and/or decrease (within
the experimental errors) the lifetime ratio $\tau(B^-)/\tau(B_d)$.
Allowing for a downward fluctuation of this ratio by two standard
deviations, i.e.\ $\tau(B^-)/\tau(B_d)>0.94$, and using the estimate
of $1/m_b^2$ corrections in (\ref{taucrude}), we conclude that
\begin{equation}\label{didef}
   {\tau(\Lambda_b)\over\tau(B_d)} > 0.97 \times \Bigg[ 0.98
   - {\Gamma_{\rm spec}(\Lambda_b)\over\Gamma(\Lambda_b)} \Bigg]
   = 0.95 - (d_1 + d_2\widetilde B)\,r \,,
\end{equation}
where $\Gamma_{\rm spec}(\Lambda_b)$ is the spectator contribution to
the width of the $\Lambda_b$ baryon. The values of the coefficients
$d_i$ are given in table~\ref{tab:4}. If we assume that $r$ and
$\widetilde B$ are of order unity, we find that the spectator
contributions yield a reduction of the lifetime of the $\Lambda_b$
baryon by a few per cent, and that $\tau(\Lambda_b)/\tau(B_d)>0.9$,
in contrast with the experimental result given in (\ref{taudata}).
If, for example, we try to push the theoretical prediction by taking
the large value $\widetilde B=1.5$ (corresponding to a violation of
the valence-quark approximation by 50\%) and choosing a low matching
scale $\mu=m_b/2$, we have to require that $r>r_{\rm min}$ with
$r_{\rm min}=3.1$, 2.2 and 1.3 for $\tau(\Lambda_b)/\tau(B_d)=0.78$,
0.83 and 0.88 (corresponding to the central experimental value and
the $1\sigma$ and $2\sigma$ fluctuations). Hence, even if we allow
for an upward fluctuation of the experimental result by two standard
deviations, we need a value of $r$ that is significantly larger than
most quark-model predictions (see the discussion in
section~\ref{subs:baryon}). Clearly, a reliable field-theoretic
calculation of the parameters $r$ and $\widetilde B$ is of great
importance to support or rule out such a possibility.

\begin{table}[htb]
\centerline{\parbox{14cm}{\caption{\label{tab:4}
Coefficients $d_i$ appearing in (\protect\ref{didef}) and
(\protect\ref{d3d4}).}}}
\vspace{0.5cm}
\centerline{\begin{tabular}{|c|cccc|}\hline\hline
\rule[-0.2cm]{0cm}{0.65cm} $\mu$ & $d_1$ & $d_2$ & $d_3$ & $d_4$ \\
\hline
$m_b/2$ & 0.016 & 0.023 & 0.178 & $-0.201$ \\
$m_b$   & 0.012 & 0.021 & 0.173 & $-0.195$ \\
$2 m_b$ & 0.008 & 0.020 & 0.167 & $-0.189$ \\
\hline\hline
\end{tabular}}
\end{table}

On the other hand, the low experimental value of the semileptonic
branching ratio may find its explanation in a low renormalization
scale (see section~\ref{sec:Bsl} below), or it may be caused by the
effects of New Physics, such as an enhanced rate for flavour-changing
neutral currents of the type $b\to s g$ \cite{Grza}--\cite{Giud}.
Hence, one may be misled in using the semileptonic branching ratio as
a constraint on the size of spectator contributions. Then there is
the possibility to decrease the value of $\tau(\Lambda_b)/\tau(B_d)$
by increasing the lifetime of the $B_d$ meson, i.e.\ in (\ref{didef})
we can allow for spectator contributions to the width of the $B_d$
meson. From table~\ref{tab:2}, it follows that the contributions of
the parameters $B_1$ and $B_2$ are very small (of order $10^{-3}$)
and can safely be neglected. Thus, we obtain
\begin{equation}\label{d3d4}
   {\tau(\Lambda_b)\over\tau(B_d)}
   \simeq 0.98 - (d_1 + d_2\widetilde B)\,r
   - (d_3\varepsilon_1 + d_4\varepsilon_2) \,,
\end{equation}
where the coefficients $d_3$ and $d_4$ are also shown in
table~\ref{tab:4}. At first sight, it seems that with a positive
$\varepsilon_1$ and a negative $\varepsilon_2$ of order $1/N_c$ one
could gain a contribution of about $-0.1$, which would take away much
of the discrepancy between theory and experiment. However, the
experimental result for the lifetime ratio $\tau(B^-)/\tau(B_d)$
imposes a useful constraint. Using (\ref{constr}) to eliminate
$\varepsilon_1$ from the relation (\ref{d3d4}), and allowing the
parameters $B_i$ to take values between 0 and 2, we find
\begin{equation}\label{LamBd}
   {\tau(\Lambda_b)\over\tau(B_d)} \simeq 0.98 \pm 0.02
   + 0.15\varepsilon_2 - (d_1 + d_2\widetilde B)\,r
   > 0.88 - (d_1 + d_2\widetilde B)\,r \,.
\end{equation}
where in the last step we have assumed that $|\varepsilon_2|<0.5$,
which we consider to be a very conservative bound. Even in this
extreme case, a significant contribution must still come from the
parameters $r$ and $\widetilde B$.

In view of the above discussion, the short $\Lambda_b$ lifetime
remains a potential problem for the heavy-quark theory. If the
current experimental value persists, there are two possibilities:
either some hadronic matrix elements of four-quark operators are
significantly larger than naive expectations based on large-$N_c$
counting rules and the quark model, or (local) quark--hadron duality,
which is assumed in the calculation of lifetimes, fails in
non-leptonic inclusive decays. In the second case, the explanation of
the puzzle lies beyond the heavy-quark expansion. Let us, therefore,
consider the first possibility and give a numerical example for some
possible scenarios. Assume that $\mu=m_b/2$ is an appropriate scale
to use in the evaluation of the Wilson coefficients, and that
$\widetilde B=1.5$. Then, to obtain $\tau(\Lambda_b)/\tau(B_d)=0.8$
without enhancing the prediction for the semileptonic branching ratio
requires $r\simeq 3$, i.e.\ several times larger than quark-model
estimates. If, on the other hand, we consider $r=1.5$ as the largest
conceivable value, we need $\varepsilon_2\simeq -0.5$, corresponding
to a rather large matrix element of the colour-octet operator $T_{\rm
S-P}$. Such a value of $\varepsilon_2$ leads to an enhancement of the
$B$-meson lifetime, and hence to an enhancement of the semileptonic
branching ratio of $B$ mesons, by $\Delta B_{\rm SL}\simeq 1\%$. As
we will discuss in section~\ref{sec:Bsl}, this is still tolerable
provided yet unknown higher-order corrections confirm the use of a
low renormalization scale. Although in both cases some large
parameters are needed, we find it important to note that until
reliable field-theoretic calculations of the matrix elements of
four-quark operators become available, a conventional explanation of
the $\Lambda_b$-lifetime puzzle cannot be excluded.

\subsection{Relation with the conventional factorization approach}
\label{sec:fact}

We now discuss in detail the relation of our approach with previous
analyses based on the factorization approximation \cite{liferef}.
This approximation amounts to setting (see appendix~A)
\begin{equation}
   B_i(\mu_{\rm had}) = \left(
   {\alpha_s(m_b)\over\alpha_s(\mu_{\rm had})} \right)^{4/\beta_0}
   \,, \qquad \varepsilon_i(\mu_{\rm had}) = 0 \,,
\end{equation}
at some hadronic scale $\mu_{\rm had}\ll m_b$. Here $\beta_0$ is the
first coefficient of the $\beta$ function. The evolution of the
operators from $m_b$ to $\mu_{\rm had}$ is done to leading
logarithmic order \cite{ShiV,hybr,PoWi}. For the purpose of our
discussion, we fix $\mu_{\rm had}$ such that $\alpha_s(\mu_{\rm
had})=0.5$. Then the relation between the hadronic parameters
renormalized at the scale $m_b$ with those renormalized at $\mu_{\rm
had}$, as given in (\ref{lowpara}), is:
\begin{eqnarray}\label{Biei}
   B_i(m_b) &\simeq& \phantom{-}1.48 B_i(\mu_{\rm had})
    - 0.36\varepsilon_i(\mu_{\rm had}) \,, \nonumber\\
   \varepsilon_i(m_b) &\simeq& -0.08 B_i(\mu_{\rm had})
    + 1.06\varepsilon_i(\mu_{\rm had}) \,.
\end{eqnarray}

The theoretical prediction for the lifetime ratio
$\tau(B^-)/\tau(B_d)$ in terms of the hadronic parameters
renormalized at the scale $\mu_{\rm had}$ is obtained by combining
(\ref{Biei}) with the numbers given in table~\ref{tab:3}. We find
\begin{equation}\label{ratlow}
   {\tau(B^-)\over\tau(B_d)} \simeq 1 + 0.08 B_1(\mu_{\rm had})
   -0.01 B_2(\mu_{\rm had}) - 0.75\varepsilon_1(\mu_{\rm had})
   + 0.20\varepsilon_2(\mu_{\rm had}) \,.
\end{equation}
The factorization approximation $B_i(\mu_{\rm had})\simeq 0.68$ and
$\varepsilon_i(\mu_{\rm had})=0$ leads to $\tau(B^-)/\tau(B_d)\simeq
1.05$, which is close to the value obtained by Bigi et al.\
\cite{liferef}. However, it is evident from (\ref{ratlow}) that even
at a low scale the coefficients of the non-factorizable terms are
still much larger than those of the factorizable ones. Hence, it
remains true that non-factorizable corrections are potentially
important. It would be justified to neglect these contributions only
if the parameters $\varepsilon_i(\mu_{\rm had})$ were significantly
smaller than 10\%. However, to the best of our knowledge there is
currently no compelling argument to support such a strong
restriction.

\section{Semileptonic branching ratio of $B$ mesons}
\label{sec:Bsl}

The semileptonic branching ratio of $B$ mesons has received
considerable attention in the past. It is defined as
\begin{equation}
   B_{\rm SL} = {\Gamma(\bar B\to X\,e\,\bar\nu)\over
   \sum_\ell \Gamma(\bar B\to X\,\ell\,\bar\nu) + \Gamma_{\rm NL}
   + \Gamma_{\rm rare}} \,,
\end{equation}
where $\Gamma_{\rm NL}$ and $\Gamma_{\rm rare}$ are the inclusive
rates for non-leptonic and rare decays, respectively. The status of
the experimental results on the semileptonic branching ratio is
controversial, as there is a discrepancy between low-energy
measurements performed at the $\Upsilon(4s)$ resonance and
high-energy measurements performed at the $Z^0$ resonance. The
average value at low energies is $B_{\rm SL}=(10.37\pm 0.30)\%$
\cite{Tomasz}, whereas high-energy measurements give $B_{\rm
SL}^{(b)}=(11.11\pm 0.23)\%$ \cite{Pascal}. The superscript $(b)$
indicates that this value refers not to the $B$ meson, but to a
mixture of $b$ hadrons (approximately 40\% $B^-$, 40\% $\bar B_d$,
12\% $B_s$, and 8\% $\Lambda_b$). Assuming that the corresponding
semileptonic width $\Gamma_{\rm SL}^{(b)}$ is close to that of the
$B$ meson,\footnote{Theoretically, this is expected to be a very good
approximation.}
we can correct for this and find $B_{\rm SL}=(\tau(B)/\tau(b))\,
B_{\rm SL}^{(b)}=(11.30\pm 0.26)\%$, where $\tau(b)=(1.57\pm
0.03)$~ps is the average lifetime corresponding to the above mixture
of $b$ hadrons \cite{Joe}. The discrepancy between the low-energy and
high-energy measurements of the semileptonic branching ratio is
therefore larger than 3 standard deviations. If we take the average
and inflate the error to account for this fact, we obtain
\begin{equation}
   B_{\rm SL} = (10.90\pm 0.46)\% \,.
\end{equation}
In understanding this result, an important aspect is charm counting,
i.e.\ the measurement of the average number $n_c$ of charm hadrons
produced per $B$ decay. Theoretically, this quantity is given by
\begin{equation}\label{ncdef}
   n_c = 1 + B(\bar B\to X_{c\bar c s'})
   - B(\bar B\to\mbox{no charm}) \,,
\end{equation}
where $B(\bar B\to X_{c\bar c s'})$ is the branching ratio for decays
into final states containing two charm quarks, and $B(\bar
B\to\mbox{no charm})\sim 0.02$ \cite{Alta,Simm,Buch} is the Standard
Model branching ratio for charmless decays. Recently, two new
measurements of the average charm content have been performed. The
CLEO Collaboration has presented the value $n_c=1.16\pm 0.05$
\cite{Tomasz,ncnew}, and the ALEPH Collaboration has reported the
result $n_c=1.23\pm 0.07$ \cite{ALEPHnc}. The average is
\begin{equation}
   n_c = 1.18\pm 0.04 \,.
\end{equation}

The naive parton model predicts that $B_{\rm SL}\simeq 15\%$ and
$n_c\simeq 1.2$; however, it has been known for some time that
perturbative corrections could change these results significantly
\cite{Alta}. With the establishment of the $1/m_Q$ expansion, the
non-perturbative corrections to the parton model could be computed,
and their effect turned out to be very small. This led Bigi et al.\
to conclude that values $B_{\rm SL}<12.5\%$ cannot be accommodated by
theory, thus giving rise to a puzzle referred to as the ``baffling
semileptonic branching ratio'' \cite{baff}. Recently, Bagan et al.\
have completed the calculation of the $O(\alpha_s)$ corrections
including the effects of the charm-quark mass \cite{Baga}, finding
that they lower the value of $B_{\rm SL}$ significantly.

\begin{figure}[htb]
   \centerline{
   \epsfysize=5cm\epsffile{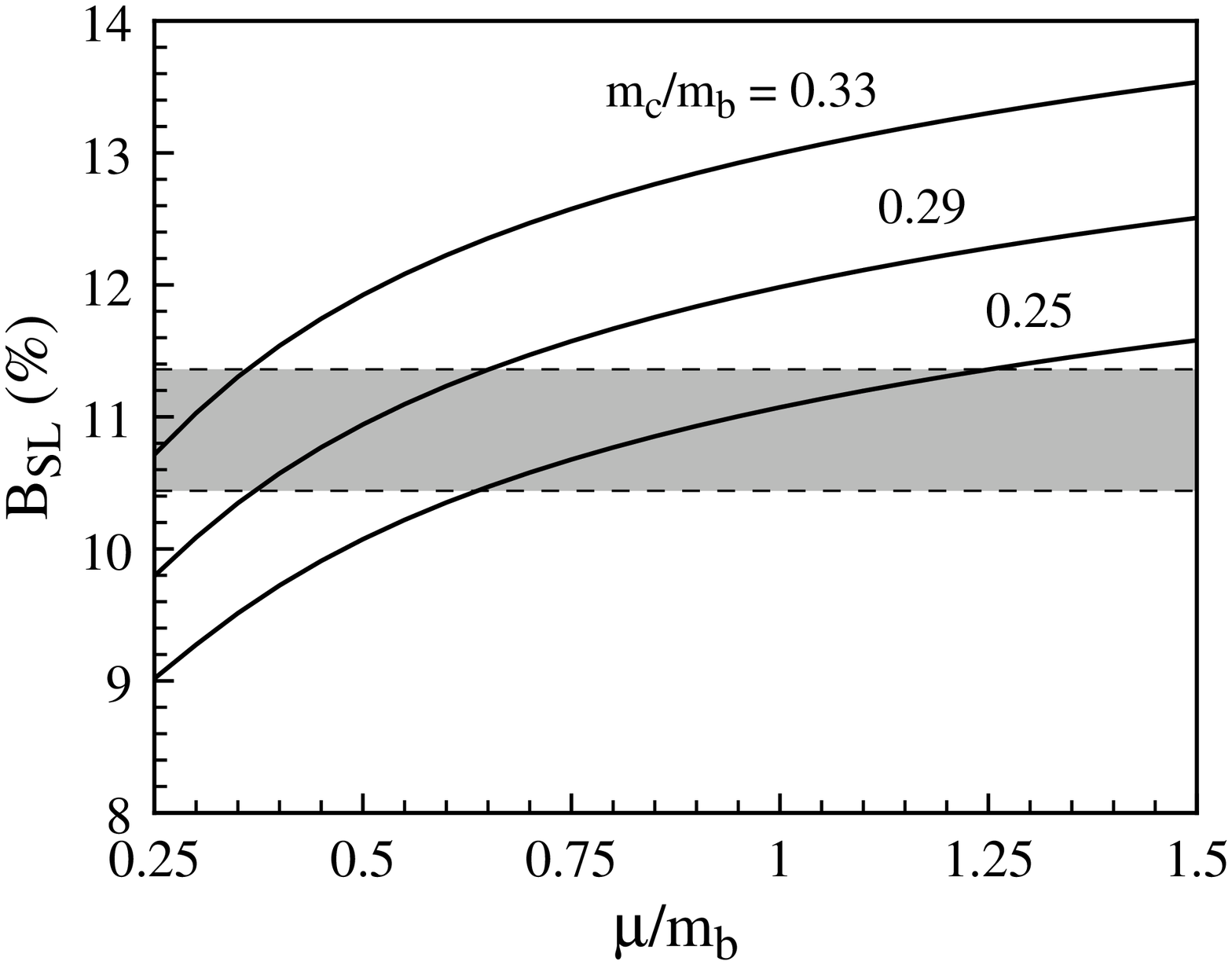}
   \epsfysize=5cm\epsffile{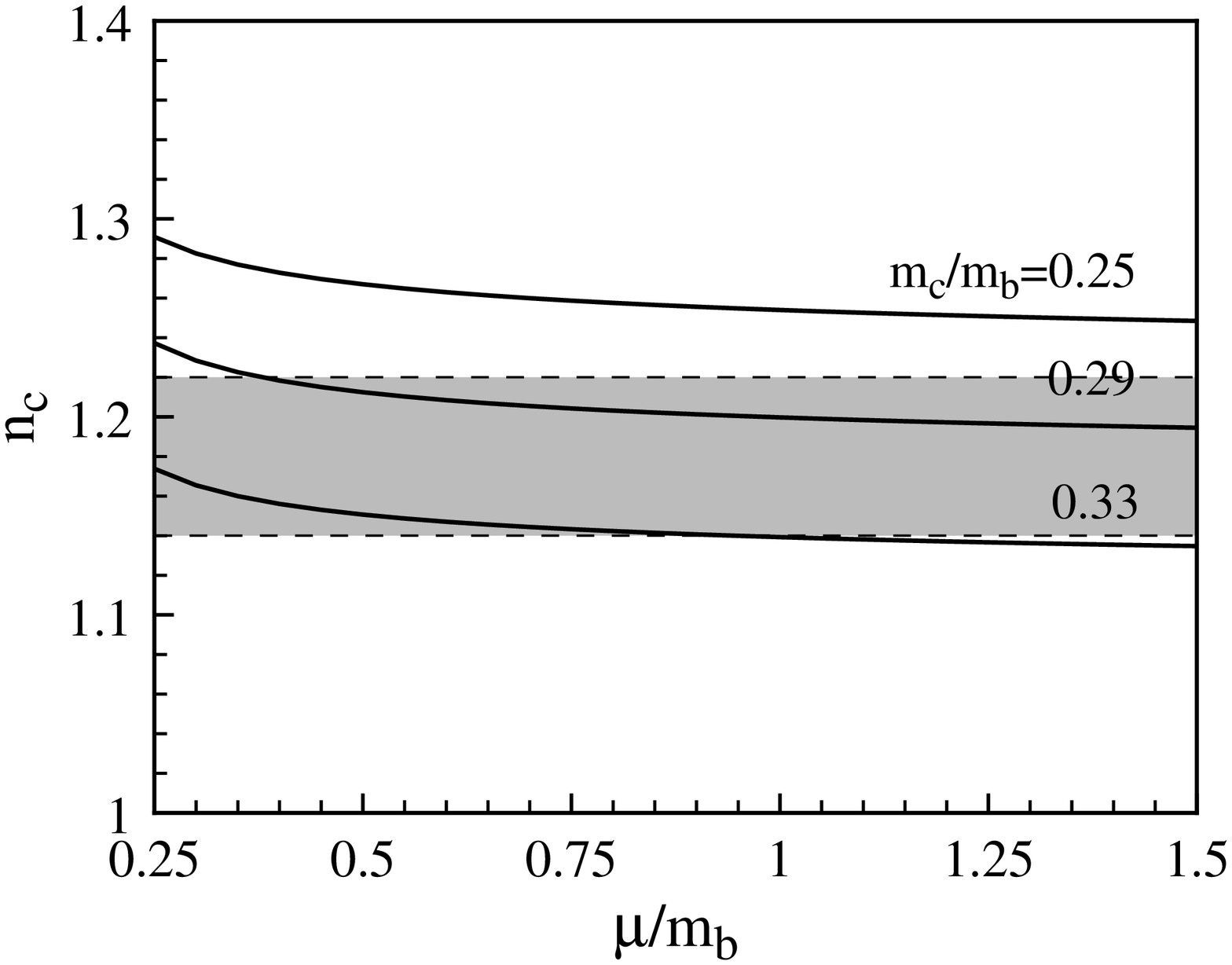}}
   \centerline{\parbox{14cm}{\caption{\label{fig:mudep}
Scale dependence of the theoretical predictions for the semileptonic
branching ratio and $n_c$. The bands show the average experimental
values.
   }}}
\end{figure}

The analysis of Bagan et al.\ has been corrected in an erratum
\cite{Baga}. Here we shall present the results of an independent
numerical analysis using the same theoretical input. As the subject
is of considerable importance, we shall explain our analysis in
detail. The semileptonic branching ratio and $n_c$ depend on the pole
masses of the heavy quarks, which we allow to vary in the range
\begin{equation}
   m_b = (4.8\pm 0.2)~\mbox{GeV} \,, \qquad
   m_b - m_c = (3.40\pm 0.06)~\mbox{GeV} \,,
\end{equation}
corresponding to $0.25< m_c/m_b <0.33$. Here $m_b$ is the pole mass
defined to one-loop order in perturbation theory. The difference
$m_b-m_c$ is free of renormalon ambiguities and can be determined
from spectroscopy (see, e.g., ref.~\cite{Beijing}). Bagan et al.\
have also considered the theoretical predictions in a scheme where
the quark masses are renormalized at a scale $\mu$ in the
$\overline{\mbox{\sc ms}}$ scheme. We discuss this scheme in
appendix~B. At order $1/m_b^2$ in the heavy-quark expansion,
non-perturbative effects are described by the single parameter
$\mu_G^2(B)$ defined in (\ref{muvals}); the dependence on the
parameter $\mu_\pi^2(B)$ is the same for all inclusive decay rates
and cancels out in $B_{\rm SL}$ and $n_c$. Moreover, the results
depend on the scale $\mu$ used to renormalize the coupling constant
$\alpha_s(\mu)$ and the Wilson coefficients $c_\pm(\mu)$ entering the
non-leptonic decay rate. They also depend on the value of the QCD
scale parameter $\Lambda_{\rm QCD}$, which we fix taking
$\alpha_s(m_Z)=0.117\pm 0.004$. The corresponding uncertainty is
smaller than that due to the variation of the mass parameters and is
added in quadrature. For the two choices $\mu=m_b$ and $\mu=m_b/2$,
we obtain
\begin{eqnarray}
   B_{\rm SL} &=& \cases{
    12.0\pm 1.0 \% ;& $\mu=m_b$, \cr
    10.9\pm 1.0 \% ;& $\mu=m_b/2$, \cr} \nonumber\\
   \phantom{ \bigg[ }
   n_c &=& \cases{
    1.20\mp 0.06 ;& $\mu=m_b$, \cr
    1.21\mp 0.06 ;& $\mu=m_b/2$. \cr}
\end{eqnarray}
The errors in the two quantities are anticorrelated. Notice that the
semileptonic branching ratio has a much stronger scale dependence
than $n_c$. This is illustrated in figure~\ref{fig:mudep}, where we
show the two quantities as a function of $\mu$. By choosing a low
renormalization scale, values $B_{\rm SL}<12.5\%$ can easily be
accommodated. The experimental data prefer a scale $\mu/m_b\sim
0.3$--0.5, which is indeed not unnatural. Using the BLM scale-setting
method \cite{BLM}, Luke et al.\ have estimated that $\mu\simeq 0.3
m_b$ is an appropriate scale in this case \cite{LSW}.

\begin{figure}[htb]
   \epsfxsize=7cm
   \centerline{\epsffile{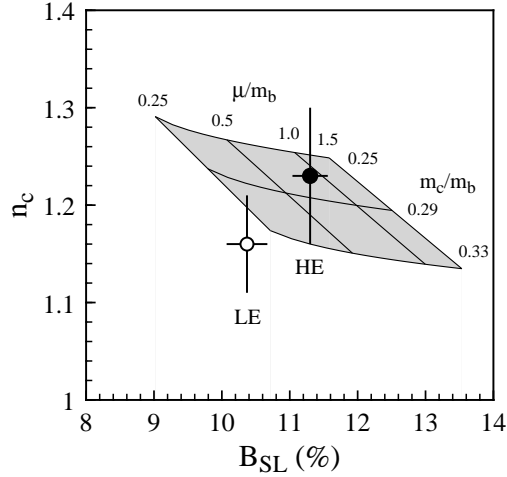}}
   \centerline{\parbox{14cm}{\caption{\label{fig:BSL}
Combined theoretical predictions for the semileptonic branching ratio
and charm counting as a function of the quark-mass ratio $m_c/m_b$
and the renormalization scale $\mu$. The data points show the average
experimental values for $B_{\rm SL}$ and $n_c$ obtained in low-energy
(LE) and high-energy (HE) measurements.
   }}}
\end{figure}

The combined theoretical predictions for the semileptonic branching
ratio and charm counting are shown in figure~\ref{fig:BSL}. They are
compared with the experimental results obtained at the $\Upsilon(4s)$
and at the $Z^0$ resonance. It was argued that the combination of a
low semileptonic branching ratio and a low value of $n_c$ would
constitute a potential problem for the Standard Model \cite{Buch}.
However, with the new experimental and theoretical numbers, only for
the low-energy measurements a small discrepancy remains between
theory and experiment. Note that, using (\ref{ncdef}), our results
for $n_c$ can be used to obtain predictions for the branching ratio
$B(\bar B\to X_{c\bar c s'})$, which is accessible to a direct
experimental determination. Our prediction of $(22\pm 6)\%$ for this
branching ratio agrees well with the preliminary result of the CLEO
Collaboration: $B(\bar B\to X_{c\bar c s'})=(23.9\pm 3.8)\%$
\cite{Hons}.

Having discussed the status of the theoretical predictions obtained
to order $1/m_b^2$ in the heavy-quark expansion, we now investigate
the spectator contributions to the semileptonic branching ratio and
$n_c$. This extends, in the context of the heavy-quark expansion, the
phenomenological study presented in ref.~\cite{Alta}. We consider the
average of $B_{\rm SL}$ and $n_c$ for $B^-$ and $B_d$
mesons,\footnote{These are approximately the quantities measured
experimentally. However, measurements at LEP receive a contamination
from $B_s$ and $b$-baryon decays.} and write for the spectator
contributions to these quantities
\begin{eqnarray}\label{bini}
   \Delta B_{\rm SL,spec}
   &=& b_1 B_1 + b_2 B_2 + b_3\varepsilon_1 + b_4\varepsilon_2 \,,
    \nonumber\\
   \Delta n_{c,{\rm spec}}
   &=& n_1 B_1 + n_2 B_2 + n_3\varepsilon_1 + n_4\varepsilon_2 \,.
\end{eqnarray}
The coefficients $b_i$ and $n_i$ are given in table~\ref{tab:5}. If,
as previously, we eliminate $\varepsilon_1$ from these equations
using the constraint (\ref{constr}) imposed by the measurement of
$\tau(B^-)/\tau(B_d)$, and we allow that the parameters $B_i$ take
values in the range 0 to 2, we obtain
\begin{eqnarray}
   \Delta B_{\rm SL,spec} &\simeq& (-2.1\varepsilon_2
    + 0.2\pm 0.3)\,\% \,, \nonumber\\
   \Delta n_{c,{\rm spec}} &\simeq& (1.2\pm 0.1)\,
    \Delta B_{\rm SL,spec} \,.
\end{eqnarray}
For reasonable value of $\varepsilon_2$, we expect a contribution to
the semileptonic branching ratio of order 1\% or less, and a
negligible effect on $n_c$. However, without a detailed calculation
of the hadronic parameters we cannot obtain a quantitative prediction
of the spectator contributions. Nevertheless, we find it interesting
that there is at least a potential to change, and in particular to
lower, the value of $B_{\rm SL}$ by 0.5--1\%. To achieve such a
decrease requires that the hadronic parameter $\varepsilon_2$, which
parametrizes the matrix element of the colour octet--octet operator
$T_{S-P}^q$ in (\ref{4qops}), is positive and of order
0.3--0.5.\footnote{We recall, however, that according to
(\protect\ref{LamBd}) a positive value of $\varepsilon_2$ increases
the theoretical prediction for the lifetime ratio
$\tau(\Lambda_b)/\tau(B_d)$.}
It will be interesting to see if future calculations of this
parameter will confirm or rule out this scenario.

\begin{table}[htb]
\centerline{\parbox{14cm}{\caption{\label{tab:5}
Coefficients $b_i$ and $n_i$ (in \%) appearing in
(\protect\ref{bini}).}}}
\vspace{0.5cm}
\centerline{\begin{tabular}{|c|cccc|}\hline\hline
\rule[-0.2cm]{0cm}{0.65cm} $\mu$ & $b_1$ & $b_2$ & $b_3$ & $b_4$ \\
\hline
$m_b/2$ & 0.35 & $-0.02$ & $-2.60$ & $-1.37$ \\
$m_b$   & 0.19 & $-0.03$ & $-2.56$ & $-1.42$ \\
$2 m_b$ & 0.03 & $-0.05$ & $-2.49$ & $-1.42$ \\
\hline\hline
\rule[-0.2cm]{0cm}{0.65cm} $\mu$ & $n_1$ & $n_2$ & $n_3$ & $n_4$ \\
\hline
$m_b/2$ & 0.50 & $-0.03$ & $-4.17$ & $-1.54$ \\
$m_b$   & 0.25 & $-0.03$ & $-3.79$ & $-1.44$ \\
$2 m_b$ & 0.03 & $-0.05$ & $-3.51$ & $-1.36$ \\
\hline\hline
\end{tabular}}
\end{table}

\section{Conclusions}
\label{sec:concs}

In this paper, we have studied spectator effects in inclusive decays
of beauty hadrons. Although these effects are suppressed by three
powers of $\Lambda_{\rm QCD}/m_b$ in the heavy-quark expansion, they
cannot be neglected because of the large phase-space factor for
two-body scattering. The contributions of spectator effects to
inclusive decay rates are given by the hadronic matrix elements of
the four local operators in (\ref{4qops}). For mesons, we have
expressed these matrix elements in terms of the hadronic parameters
$B_{1,2}$ and $\varepsilon_{1,2}$ defined in (\ref{Bidef}). For the
$\Lambda_b$ baryon, heavy-quark symmetry reduces the number of
independent matrix elements from four to two, which we parametrize by
$r$ and $\widetilde B$ as defined in (\ref{Btildef}) and
(\ref{rdef}). Although our parametrization is motivated by commonly
made simplifications, such as the vacuum insertion and the
valence-quark approximations, we stress that it is introduced without
any loss of generality. For a complete understanding of spectator
effects, it will be necessary to evaluate these parameters
non-perturbatively, e.g.\ in lattice simulations.

We find that in predictions for lifetimes and the semileptonic
branching ratio of $B$ mesons, the coefficients of the colour
octet--octet non-factorizable operators are much larger than those
for the colour singlet--singlet factorizable operators. Thus the
contributions from the non-factorizable operators cannot be
neglected, even though their matrix elements are suppressed in the
large-$N_c$ limit.

The ratio $\tau(B^-)/\tau(B_d)$ is particularly sensitive to
non-factorizable contributions [see (\ref{cidef}) and
table~\ref{tab:3}], making it difficult to predict this quantity with
a precision of better than about 10\%. For example, assuming that the
magnitudes of the parameters $\varepsilon_1$ and $\varepsilon_2$ are
smaller than 0.1 or 0.2, we find that the predictions for this
lifetime ratio lie in the ranges 0.93--1.11 and 0.84--1.20,
respectively.\footnote{Alternatively, if we assume that the
magnitudes of $\varepsilon_1(\mu_{\rm had})$ and
$\varepsilon_2(\mu_{\rm had})$ renormalized at a hadronic scale are
less than 0.1 and 0.2, then, using (\ref{ratlow}), we find that the
corresponding predictions lie in the ranges 0.96--1.14 and
0.87--1.23.}
However, in our opinion, even if the experimental result had been
outside these ranges, the most likely explanation would have been
that the $\varepsilon_i$ parameters are larger, rather than a failure
of the heavy-quark expansion. The experimental measurement of
$\tau(B^-)/\tau(B_d)$ imposes the constraint (\ref{constr}) upon the
parameters, which allows us to eliminate $\varepsilon_1$ in other
relations. On the other hand, within the heavy-quark expansion there
is only room for a very small deviation of the ratio
$\tau(B_s)/\tau(B_d)$ from unity due to SU(3)-breaking effects. We
estimate these effects to be of order 1\%.

Understanding the low experimental value of the lifetime ratio
$\tau(\Lambda_b)/\tau(B_d)$ remains a potential problem for the
heavy-quark theory. If the current experimental value persists, there
are two possibilities: either some hadronic matrix elements of
four-quark operators are significantly larger than naive expectations
based on large-$N_c$ counting rules and the quark model, or (local)
quark--hadron duality fails in non-leptonic inclusive decays. In the
second case, the explanation of the puzzle lies beyond the
heavy-quark expansion. In the first case, it is most likely that the
baryonic parameter $r$ is much larger than most expectations based on
quark-model estimates. It will be interesting to see whether future,
field-theoretic calculations will yield values of $r$ which are
sufficiently large. Until such calculations become available, a
conventional explanation of the $\Lambda_b$-lifetime puzzle cannot be
excluded.

Finally, we have performed an analysis of the semileptonic branching
ratio of the $B$ meson ($B_{\rm SL}$) and of the average number of
charmed particles produced per decay ($n_c$). Our results are
summarized in figures~\ref{fig:mudep} and \ref{fig:BSL}. There is a
significant dependence on the predictions for the semileptonic
branching ratio on the renormalization scale $\mu$, which is a
manifestation of our ignorance of higher-order perturbative
corrections. The results for $n_c$, on the other hand, are almost
independent of $\mu$. This scale dependence weakens considerably the
anticorrelation in the theoretically allowed values for $B_{\rm SL}$
and $n_c$ observed in ref.~\cite{Buch}. In our view, given the
theoretical uncertainties and the disagreement between the
experimental values for the semileptonic branching ratio obtained in
low- and high-energy measurements, there is at present no discrepancy
between theory and experiment for $B_{\rm SL}$ and $n_c$. We have
also studied the contributions of spectator effects for these
quantities and find that they are negligible for $n_c$, whereas they
can potentially change the prediction for $B_{\rm SL}$ by up to about
1\%.

\vspace{0.3cm}
{\it Note added:\/}
After completing this work we became aware of a paper by I.I.~Bigi
(preprint UND-HEP-96-BIG01, June 1996 [hep-ph/9606405]), who
discusses theoretical predictions for beauty lifetimes, making strong
claims concerning the theoretical predictions for the lifetime ratio
$\tau(B^-)/\tau(B_d)$. In view of our discussion in
section~\ref{sec:fact}, we must disagree with some statements made in
this paper.

\vspace{0.3cm}
{\it Acknowledgements:\/}
We thank Patricia Ball, Jon Rosner, Berthold Stech and Kolia Uraltsev
for helpful discussions. C.T.S.\ acknowledges the Particle Physics
and Astronomy Research Council for its support through the award of a
Senior Fellowship.

\newpage
\setcounter{equation}{0}
\renewcommand{\thesection}{Appendix~A:}
\renewcommand{\theequation}{A.\arabic{equation}}

\section{Renormalization-group evolution}

The four-quark operators appearing in the heavy-quark expansion are
conventionally renormalized at the scale $\mu=m_b$. However, one may
use the renormalization-group to rewrite them in terms of operators
renormalized at a scale $\mu\ne m_b$. The renormalization-group
evolution is determined by the anomalous dimensions of the four-quark
operators in the HQET, where the $b$ quark is treated as static quark
\cite{review}. In the literature, this evolution is sometimes
referred to as ``hybrid renormalization'' \cite{ShiV,hybr,PoWi}.

We find that the operators $O_{V-A}^q$ and $T_{V-A}^q$, and similarly
$O_{S-P}^q$ and $T_{S-P}^q$, mix under renormalization. At one-loop
order, the mixing within each pair $(O,T)$ is governed by the
anomalous dimension matrix
\begin{equation}
   \hat\gamma = {3\alpha_s\over 2\pi}\,\left(
   \begin{array} {cc} \phantom{ \bigg[ } C_F & 1 \\
    \displaystyle -{C_F\over 2 N_c}~ & \displaystyle ~{1\over 2 N_c}
   \end{array} \right) \,,
\end{equation}
which has eigenvalues 0 and $3 N_c$. Here $N_c$ is the number of
colours, and $C_F = (N_c^2-1)/2 N_c$ is the eigenvalue of the
quadratic Casimir operator in the fundamental representation. The
operators defined at the scale $m_b$ can be rewritten in terms of
those defined at a scale $\mu\ne m_b$. In leading logarithmic
approximation, the result is
\begin{eqnarray}\label{evol}
   O(m_b) &=& \bigg[ 1 + {2 C_F\over N_c}\,(\kappa-1) \bigg]\,
    O(\mu) - {2\over N_c}\,(\kappa-1)\,T(\mu) \,, \nonumber\\
   T(m_b) &=& \bigg[ 1 + {1\over N_c^2}\,(\kappa-1) \bigg]\,T(\mu)
    - {C_F\over N_c^2}\,(\kappa-1)\,O(\mu) \,,
\end{eqnarray}
where
\begin{equation}
   \kappa = \left( {\alpha_s(\mu)\over\alpha_s(m_b)}
   \right)^{3 N_c/2\beta_0} \,,
\end{equation}
and $\beta_0=\frac{11}{3}\,N_c-\frac{2}{3}\,n_f$ is the first
coefficient of the $\beta$-function ($n_f=3$ is the number of light
quark flavours).

Given the evolution equations (\ref{evol}) for the four-quark
operators, it is immediate to derive the corresponding equations for
the hadronic parameters defined in (\ref{Bidef}), (\ref{Btildef}) and
(\ref{rdef}). We obtain:
\begin{eqnarray}\label{evolve}
   B_i(m_b) &=& \bigg[ 1 + {2 C_F\over N_c}\,(\kappa-1) \bigg]\,
    B_i(\mu) - {2\over N_c}\,(\kappa-1)\,\varepsilon_i(\mu)
    \,, \nonumber\\
   \varepsilon_i(m_b) &=& \bigg[ 1 + {1\over N_c^2}\,(\kappa-1)
    \bigg]\,\varepsilon_i(\mu) - {C_F\over N_c^2}\,(\kappa-1)\,
    B_i(\mu) \,, \nonumber\\
   r(m_b) &=& \kappa\,r(\mu) + {1\over N_c}\,(\kappa-1)\,
    \widetilde B(\mu)\,r(\mu) \,, \nonumber\\
   \widetilde B(m_b)\,r(m_b) &=& \widetilde B(\mu)\,r(\mu) \,.
\end{eqnarray}
Of course, introducing parameters renormalized at a scale $\mu\ne
m_b$ would simply amount to a reparametrization of the results and as
such is not very illuminating. However, the evolution equations are
used in section~\ref{sec:param} to study the sensitivity of our
results to unknown higher-order corrections.

As an illustration, we study the evolution from $\mu=m_b$ down to a
typical hadronic scale $\mu_{\rm had}\ll m_b$, which we choose such
that $\alpha_s(\mu_{\rm had})=0.5$ (corresponding to $\mu_{\rm
had}\sim 0.67$~GeV). We find
\begin{eqnarray}\label{lowpara}
   B_i(m_b) &\simeq& 1.48 B_i(\mu_{\rm had})
    - 0.36\varepsilon_i(\mu_{\rm had}) \,, \nonumber\\
   \varepsilon_i(m_b) &\simeq& 1.06\varepsilon_i(\mu_{\rm had})
    - 0.08 B_i(\mu_{\rm had}) \,, \nonumber\\
   r(m_b) &\simeq& \big[ 1.54 + 0.18\widetilde B(\mu_{\rm had})
    \big]\,r(\mu_{\rm had}) \,, \nonumber\\
   \widetilde B(m_b) &\simeq& {\widetilde B(\mu_{\rm had})\over
    1.54 + 0.18\widetilde B(\mu_{\rm had})} \,,
\end{eqnarray}
indicating that renormalization effects can be quite significant. If
one assumes that the matrix elements renormalized at the scale
$\mu_{\rm had}$ can be estimated using the vacuum insertion
hypothesis for mesons and the valence-quark approximation for
baryons, then
\begin{equation}\label{muhad}
   B_i(\mu_{\rm had})\simeq {f_B^2(\mu_{\rm had})\over f_B^2(m_b)}
   = \kappa^{-8/9} \simeq 0.68 \,, \qquad
   \varepsilon_i(\mu_{\rm had})\simeq 0 \,, \qquad
   \widetilde B(\mu_{\rm had})\simeq 1 \,,
\end{equation}
where the factor $\kappa^{-8/9}$ in the first equation arises from
the anomalous dimension of the axial current in the HQET
\cite{hybr,PoWi}. Using (\ref{evolve}), we then find that the
parameters defined at a renormalization scale $m_b$ (as used
throughout this paper) would be $B_1(m_b)=B_2(m_b)\simeq 1.01$,
$\varepsilon_1(m_b)= \varepsilon_2(m_b)\simeq -0.05$,
$r(m_b)/r(\mu_{\rm had})\simeq 1.72$, and $\widetilde B(m_b)\simeq
0.58$. Thus, for mesons the violations of the factorization
approximation induced by the evolution from $\mu_{\rm had}$ up to
$m_b$ remain small, i.e.\ we find $B_i(m_b)\simeq 1$ and
$\varepsilon_i(m_b)\simeq 0$. We stress, however, that we do not want
to suggest that the choice of parameters in (\ref{muhad}) is actually
physical.

For mesons, the large-$N_c$ counting rules \cite{Witt,BGR}
imply that the parameters $\varepsilon_i$ are of order $1/N_c$,
whereas the parameters $B_i$ are of order unity. These results are
respected by the evolution equations (\ref{evolve}), which in the
large-$N_c$ limit take the form
\begin{eqnarray}
   B_i(m_b) &=& \kappa_\infty\,B_i(\mu) + O(1/N_c) \,, \nonumber\\
   \varepsilon_i(m_b) &=& \varepsilon_i(\mu) - {1\over 2 N_c}\,
    (\kappa_\infty - 1)\,B_i(\mu) + O(1/N_c^2) \,,
\end{eqnarray}
where $\kappa_\infty=[\alpha_s(\mu)/\alpha_s(m_b)]^{9/22}$. Under a
change of the renormalization scale, the parameters $\varepsilon_i$
stay of order $1/N_c$, whereas the parameters $B_i$ change by a
factor of order unity.

\setcounter{equation}{0}
\renewcommand{\thesection}{Appendix~B:}
\renewcommand{\theequation}{B.\arabic{equation}}

\section{$B_{\rm SL}$ and $n_c$ in the $\overline{\mbox{\sc ms}}$
scheme}

The semileptonic branching ratio and $n_c$ can also be calculated
using running quark masses renormalized in the $\overline{\mbox{\sc
ms}}$ scheme rather than pole masses. To compare the results in such
a scheme to those presented in our work, we have to relate the ratio
of the pole masses to the ratio of the running masses. There is some
freedom in how to do this translation. Since in the expressions for
the (partial) inclusive decay rates radiative corrections are
included to order $\alpha_s(\mu)$ only, it is consistent to work with
the one-loop relation
\begin{equation}
   {m_c\over m_b} = {\overline{m}_c(\mu)\over\overline{m}_b(\mu)}\,
   \Bigg( 1 - {2\alpha_s(\mu)\over\pi}\,
   \ln{\overline{m}_c(\mu)\over\overline{m}_b(\mu)} \Bigg) \,.
\end{equation}
We shall refer to this choice as scheme $\overline{\mbox{\sc ms}}1$.
Alternatively, one may prefer to resum the leading and
next-to-leading logarithms to this relation, which leads to
\begin{equation}
   {\overline{m}_c(\mu)\over\overline{m}_b(\mu)} = {m_c\over m_b}\,
   \Bigg( {\alpha_s(m_c)\over\alpha_s(m_b)}
   \Bigg)^{\gamma_0/2\beta_0}\,\Bigg\{ 1
   - {\alpha_s(m_c) - \alpha_s(m_b)\over\pi}\,\bigg(
   {\gamma_1\beta_0 - \gamma_0\beta_1\over 8\beta_0^2}
   - {4\over 3} \bigg) \Bigg\} \,,
\end{equation}
where $\gamma_0$ and $\gamma_1$ are the one- and two-loop
coefficients of the anomalous dimension of the running quark mass,
and $\beta_0$ and $\beta_1$ are the coefficients of the
$\beta$-function. We shall call this scheme $\overline{\mbox{\sc
ms}}2$; it has been adopted in the work of Bagan et al.~\cite{Baga}.

\begin{figure}[htb]
   \epsfxsize=7cm
   \centerline{\epsffile{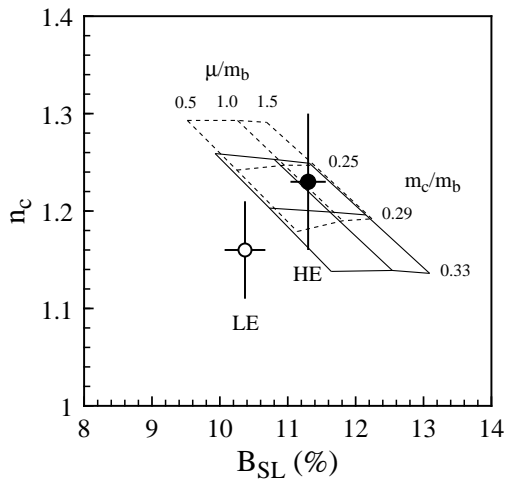}}
   \centerline{\parbox{14cm}{\caption{\label{fig:BSLMS}
Combined theoretical predictions for the semileptonic branching ratio
and charm counting as a function of the quark-mass ratio $m_c/m_b$
and the renormalization scale $\mu$. The solid lines refer to the
scheme $\overline{\mbox{\sc ms}}1$, the dashed ones to
$\overline{\mbox{\sc ms}}2$.
}}}
\end{figure}

Our results for these two versions of the $\overline{\mbox{\sc ms}}$
scheme are:
\begin{eqnarray}
   B_{\rm SL}(\overline{\mbox{\sc ms}}1) &=& \cases{
    11.6\pm 0.9 \% ;& $\mu=m_b$, \cr
    10.7\pm 0.9 \% ;& $\mu=m_b/2$, \cr} \nonumber\\
   \phantom{ \bigg[ }
   n_c(\overline{\mbox{\sc ms}}1) &=& \cases{
    1.20\mp 0.06 ;& $\mu=m_b$, \cr
    1.20\mp 0.06 ;& $\mu=m_b/2$, \cr} \nonumber\\
   \\
   B_{\rm SL}(\overline{\mbox{\sc ms}}2) &=& \cases{
    10.9\pm 0.9 \% ;& $\mu=m_b$, \cr
    10.3\pm 0.9 \% ;& $\mu=m_b/2$, \cr} \nonumber\\
   \phantom{ \bigg[ }
   n_c(\overline{\mbox{\sc ms}}2) &=& \cases{
    1.25\mp 0.05 ;& $\mu=m_b$, \cr
    1.24\mp 0.06 ;& $\mu=m_b/2$, \cr} \nonumber
\end{eqnarray}
and the combined predictions for $B_{\rm SL}$ and $n_c$ are shown in
figure~\ref{fig:BSLMS}. Contrary to the case of the on-shell scheme,
the calculations in the $\overline{\mbox{\sc ms}}$ scheme become
unstable for low values of the renormalization scale. For this
reason, we only present result for $\mu\ge m_b/2$. The results
obtained in the scheme $\overline{\mbox{\sc ms}}1$ are close to those
obtained in the on-shell scheme and presented in
section~\ref{sec:Bsl}. In the scheme $\overline{\mbox{\sc ms}}2$, on
the other hand, we find lower values for $B_{\rm SL}$ and higher
values for $n_c$. We note that our results for this scheme do not
coincide with the numbers presented in the erratum of
ref.~\cite{Baga}; in particular, we do not find the large values of
$n_c$ reported there. The numerical differences are mainly due to the
fact that Bagan et al.\ use lower values for the charm-quark mass
(they use $m_c=1.33$~GeV for the central value of the pole mass
rather than 1.4~GeV), and that they multiply each partial decay rate
by the numerical factor
\begin{equation}
   \bigg( {m_b\over\overline{m}_b(\mu)} \bigg)^5
   = 1 + {\alpha_s(\mu)\over\pi}\,\bigg( {20\over 3}
   - 5\ln{m_b^2\over\mu^2} \bigg) + \dots \,,
\end{equation}
which we omit since it cancels trivially in the dimensionless
quantities $B_{\rm SL}$ and $n_c$. Note that, in particular, this
factor is responsible for the large apparent scale dependence of the
results presented in ref.~\cite{Baga}. Another difference is that we
include in the calculation an estimate of the contribution from
charmless decays, using $B_{\rm no~charm}=(2\pm 1)\%$
\cite{Alta,Simm,Buch} and $\Gamma_{\rm SL}^{b\to u}/\Gamma_{\rm
SL}^{b\to c}\simeq 1\%$. This lowers $B_{\rm SL}$ by a factor 0.99
and $n_c$ by a factor 0.98.

It may be argued that the apparent large scheme dependence of the
results for the semileptonic branching ratio and $n_c$ prevent a
reliable theoretical prediction. However, as we have shown above the
main reason is that the numerical value of the quark mass ratio
$m_c/m_b$ can be quite different in different schemes
($\overline{m}_c(\mu)/\overline{m}_b(\mu)\simeq 0.8\,m_c/m_b$ in the
scheme $\overline{\mbox{\sc ms}}1$, and $0.7\,m_c/m_b$ in
$\overline{\mbox{\sc ms}}2$). Since the dependence of $B_{\rm SL}$
and $n_c$ on the quark-mass ratio comes simply from phase space (and
is particularly strong for the channel $b\to c\bar c s$), we feel
that the on-shell scheme is more adequate for performing the
calculation. In other words, we expect that in the
$\overline{\mbox{\sc ms}}$ scheme one would encounter larger
higher-order corrections, once the calculation is pushed to order
$\alpha_s^2$ and higher.

\end{document}